\documentclass[%
pop
amsmath,amssymb,
reprint,%
author-year,%
superscriptaddress,
raggedbottom
]{revtex4-1}
\setcitestyle{super,comma}
\usepackage{hyperref}
\hypersetup{
    colorlinks=true,
    allcolors=blue,
}
\usepackage{graphicx}
\usepackage{dcolumn}
\usepackage{bm}

\newcommand{\ExB}{$\bm{E} \times \bm{B}$}
\begin{document}

\title{Gyrokinetic continuum simulations of plasma turbulence in the Texas Helimak}
\author{T.N. Bernard}
\email[]{tnbernard@utexas.edu}
\affiliation{Institute for Fusion Studies, University of Texas at Austin, Austin, TX, 78712, USA}
\author{E.L. Shi}
\affiliation{Lawrence Livermore National Laboratory, Livermore, California 94550, USA}
\affiliation{Department of Astrophysical Sciences, Princeton University, Princeton, New Jersey 08544, USA}
\author{K. Gentle}
\affiliation{Institute for Fusion Studies, University of Texas at Austin, Austin, TX, 78712, USA}
\author{A. Hakim}
\affiliation{Princeton Plasma Physics Laboratory, Princeton, New Jersey 08540, USA}
\author{G.W. Hammett}
\affiliation{Princeton Plasma Physics Laboratory, Princeton, New Jersey 08540, USA}
\author{T. Stoltzfus-Dueck}
\affiliation{Princeton Plasma Physics Laboratory, Princeton, New Jersey 08540, USA}
\author{E.I. Taylor}
\affiliation{Institute for Fusion Studies, University of Texas at Austin, Austin, TX, 78712, USA}

\begin{abstract}
The first gyrokinetic simulations of plasma turbulence in the Texas Helimak device, a simple magnetized torus, are presented. The device has features similar to the scrape-off layer region of tokamaks, such as bad-curvature-driven instabilities and sheath boundary conditions on the end plates, which are included in these simulations. Comparisons between simulations and measurements from the experiment show similarities, including equilibrium profiles and fluctuation amplitudes that approach experimental values, but also some important quantitative differences.  Both experimental and simulation results exhibit turbulence statistics that are characteristic of blob transport. 
\end{abstract}
\maketitle
\section{Introduction}
\begin{figure}
    \centering
    \includegraphics[width=0.4\textwidth]{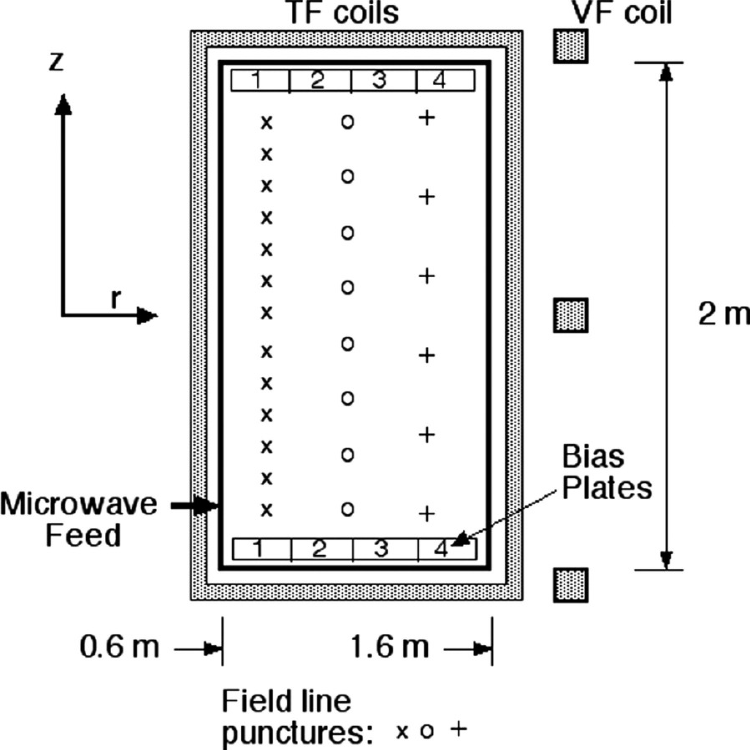}
    \caption{A cross-section of the Helimak experiment in the $(R,\varphi)$ plane. Reproduced from Gentle, K. W., et al. ``Turbulence in the cylindrical slab." Physics of Plasmas 21.9 (2014): 092302., with the permission of AIP Publishing.}
    \label{fig:hel-cross}
\end{figure}

In magnetically confined fusion devices such as tokamaks, the scrape-off layer (SOL) lies outside the last closed magnetic flux surface and consists of open magnetic field lines that intersect material walls. This region is characterized by fast transport along field lines toward the divertor plates or limiters and turbulent cross-field transport across field lines to the wall. Based on current experiments, the SOL heat-flux width is expected to be very narrow (only several gyroradii) and may not scale with machine size.\cite{Eich2013,Goldston:2012} Increasing turbulence could mitigate this by diffusing the power over a wider area. Also, turbulent coherent structures called blobs (also referred to as filaments) \cite{krasheninnikov2001scrape,krasheninnikov2008recent,zweben1985structure,zweben1985search,zweben2003high,terry2007scrape,boedo2014edge} with higher temperatures and densities than the background plasma, are convected across field lines in the SOL and can collide with chamber walls before reaching the divertor plate, potentially damaging plasma-facing components and injecting impurities into the core plasma.\cite{stangeby2000plasma} Thus, the balance between parallel and cross-field transport in the SOL plays an important role in determining how heat and particles are exhausted in advanced fusion devices, \cite{Shimada2007,loarte2007power} and accurate modeling of this region is essential to predict the performance of future fusion reactors such as ITER. \\
\indent The SOL in tokamaks is generally characterized by steep temperature and density gradients, geometry that includes an X-point, and complex physics, such as plasma-wall interactions, impurity transport, radiation, neutral recycling, and a plasma sheath. Basic plasma physics experiments with simpler magnetic configurations and lower temperature plasmas than tokamaks permit more comprehensive probe diagnostics and wider parameter scans than is generally available in fusion devices and are, thus, useful for validating analytical and numerical models of the SOL. For example, simple magnetized torus (SMT) experiments, such as the Texas Helimak\cite{gentle2008texas,perez2006drift} and TORPEX,\cite{poli2006experimental,poli2008transition,ricci2008high} are experimental approximations to a sheared slab geometry with bad curvature. They use vertical and toroidal field coils to create open, helical magnetic-field-line configurations with curvature and shear. Though SMT's do not contain the complicated geometry of a tokamak SOL, such as an X-point, the dimensionless parameters (see table 2.3 in Ref. \citenum{Williams2017thesis}) and helical, open field lines are similar to that of a tokamak SOL (Fig.~\ref{fig:hel-cross}). Thus, experimental data from these devices can be compared with results from analytic and numerical models of SOL turbulence, as has been done with various fluid codes. \cite{li2009plasma,li2011turbulence,ricci2008high,ricci2009transport,ricci2010turbulence} \\
\indent The main turbulent drives in the Helimak are the interchange and drift wave instabilities.\cite{gentle2008texas,gentle2010comparison,gentle2014turbulence,perez2006drift,Williams2017thesis} Li \emph{et al.}~developed Helimak simulations to study these instabilities using an electrostatic drift-reduced Braginskii model with cold ions.\cite{li2009plasma,li2011turbulence} These fluid models reproduced some key experimental features, and more complete fluid models may explain SMT and tokamak-SOL plasmas well in some collisional parameter regimes. However, kinetic models are required to capture effects that can be significant in lower-collisionality parameter regimes, such as trapped particles, some non-linear wave--particle interactions, and non-Maxwellian features in the particle distribution functions.\cite{cohen2008progress,scott2003computation,scott2010nonlinear} This paper details the first kinetic simulations of the Helimak. We use the Gkeyll computational framework to solve a five-dimensional (three spatial dimensions and two velocity dimensions) gyrokinetic equation, building on work by Shi \emph{et al.}~using straight and helical open-field-line configurations.\cite{shi2015gyrokinetic,shi2017gyrokinetic,Shi2017thesis,shi2019full} We maintain similarities with the set-up of the fluid simulations in Refs. \citenum{li2009plasma} and \citenum{li2011turbulence} for comparison with those results. \\
\indent In Refs. \citenum{Shi2017thesis} and \citenum{shi2019full}, Gkeyll simulations were performed with SMT geometry and NSTX-SOL parameters for a thin radial region ($\Delta R/R \sim 0.1$). These simulations neglected geometrical variation of SOL flux surfaces with poloidal angle, such as flux expansion and magnetic shear. Therefore, detailed comparisons with experimental data were not included, though there were interesting qualitative similarities. Here we use Gkeyll to model the Helimak, where the geometry of the simulation is much closer to that in the experiment, which only has a toroidal magnetic field plus a vertical magnetic field. The experiment has some magnetic shear\cite{gentle2008texas,Williams2017thesis} that is ignored at present in the simulation. The Helimak simulation also spans a much broader range of radius, $\Delta R/R \sim 0.5$. We include detailed comparisons with experimental measurements, including the time-averaged density and temperature profiles, turbulence fluctuation amplitudes, and correlation functions.\\
\indent The ratio of the vertical magnetic field to toroidal magnetic field $B_Z/B_\varphi$ can be varied experimentally from 0.16 to 0.003 by changing resistance in the vertical field coils.\cite{gentle2014turbulence} The transition between various turbulence regimes in SMT's as $B_Z/B_\varphi$ is varied has been studied numerically and experimentally.\cite{ricci2009transport,ricci2010turbulence,poli2006experimental,poli2008transition} In Ref.~\citenum{ricci2010turbulence}, Ricci and Rogers used the electrostatic, drift-reduced Braginskii equations\cite{zeiler1997nonlinear} to analyze the transition between different turbulence regimes, including ideal interchange, resistive-interchange, and drift-interchange, in SMT's. They predicted only the first and last are present in the Helimak. For the ideal interchange instability, $k_\parallel=0$, and the dispersion relation predicts a growth rate of $\gamma_I \sim c_s/\sqrt{R L_p}$, where $c_s$ is the ion sound speed, $R$ is the major radius of the SMT, and $L_p$ is the pressure gradient scale length. In the high-pitch-angle configurations of the Helimak that we consider in the present paper, the experimentally measured parallel wavelength is typically much larger than the magnetic-field-line connection length, resulting in $k_\parallel \simeq 0$.\cite{Williams2017thesis} Hence, the ideal interchange instability is expected to dominate. In the low-pitch-angle configuration, with finite $k_\parallel$, the drift-interchange instability is the main turbulent drive.

The plasma response to sheath physics in the Helimak sets up a radially-varying electric potential, producing a vertical \ExB\ flow that varies with radius. A bias voltage can be applied to conducting plates on which magnetic field lines terminate to modify the vertical sheared flow and study the effect of velocity shear on turbulence suppression. In this paper, we only consider the case where all conducting plates are grounded and present the results of limiter-biasing simulations in a future paper. \\
\indent This paper is organized as follows. In Sec.~\ref{sec:model}, we review the model equations and parameters we use in our simulations. In Sec.~\ref{sec:results}, we describe the results of our non-linear simulation, make comparisons with experimental data, identify the dominant instability, and briefly discuss kinetic effects. In Sec.~\ref{sec:conclusion}, we summarize our findings and describe future directions of the code development to improve comparisons with experimental data.
\begin{figure*}
    \centering
    \includegraphics[width=0.9\textwidth]{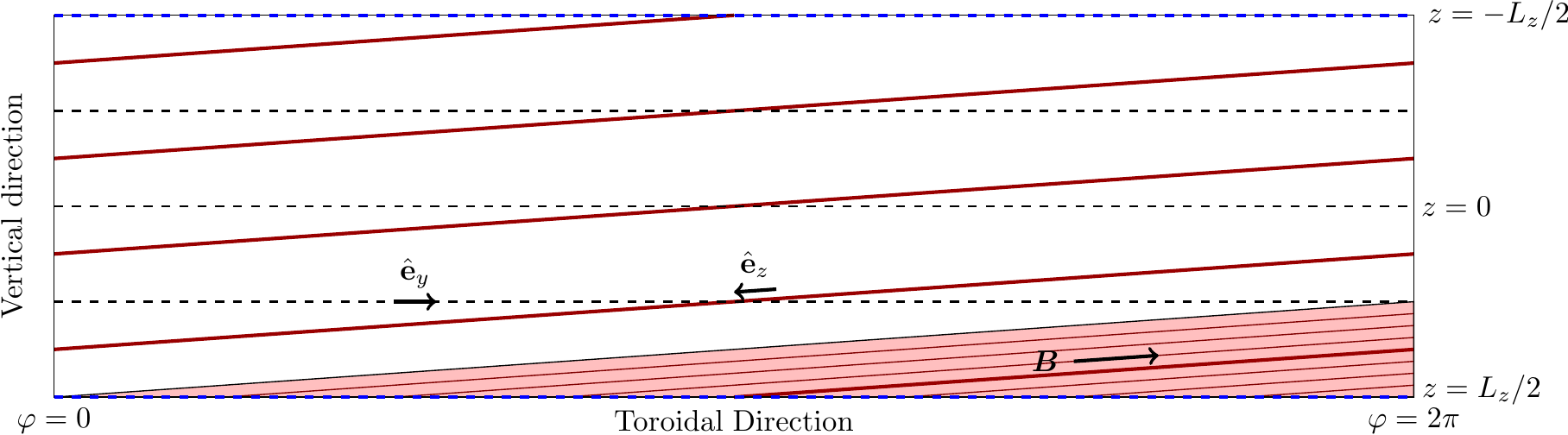}
    \caption{A map of the non-orthogonal field-line-following coordinates to cylindrical coordinates $(r,\varphi,Z)$ at a slice in the radial direction for a field line with $N=4$ toroidal turns. (The experiment we simulated actually has $N=5.8$ turns.) The pink region shows one of the edges of the simulation domain, at $z=L_z/2$. Red lines are surfaces of constant $y$, which follow field lines; i.e.~$y$ is constant on a field line. As such, it is a field line label and is used as a perpendicular coordinate. The thick red line is parallel to the magnetic field line at $y=0$. Dashed black lines are surfaces of constant $z$, with conducting sheath boundary conditions applied along the blue dashed lines, and fewer are shown to indicate the coarser resolution in $z$ than in $y$. Note that the Helimak experimental magnetic field $\bm{B}$ points in the direction of decreasing $z$, and the unit tangent vectors form a left-handed coordinate system. The radial tangent vector $\hat{\bf{e}}_x$ points out of the plane and $(\hat{\bf{e}}_x \times \hat{\bf{e}}_y) \cdot \hat{\bf{e}}_z < 0$. }
    \label{fig:flux-mapping}
\end{figure*}

\section{Model Equations} \label{sec:model}
\indent The Gkeyll code uses a nodal discontinuous Galerkin computational method for the spatial discretization and an explicit third-order Runge--Kutta method to discretize in time.\cite{liu2000high,gottlieb2001strong}  In our simulations, we solve the full-$f$ gyrokinetic equation in the long-wavelength, zero-Larmor-radius limit using the gyrocenter distribution function $f_s(\bm{R}, v_\parallel, \mu,t)$:
\begin{eqnarray}
\label{eq:gk-eqn}
\frac{\partial \mathcal{J}_s f_s}{\partial t} &+& \nabla \cdot (\mathcal{J}_s \dot{\bm{R}} f_s) + \frac{\partial}{\partial v_\parallel} (\mathcal{J}_s \dot{v_\parallel} f_s ) \nonumber \\ 
&=&\mathcal{J}_s C[f_s] + \mathcal{J}_s S_s, 
\end{eqnarray}
where $s$ refers to the species, $C[f_s]$ is a simplified nonlinear Fokker-Planck collision operator, also called the Lenard-Bernstein or Dougherty operator, and $S_s$ is a source term, which we describe in more detail below. The Jacobian is $\mathcal{J} = B_\parallel^*$, where $B_\parallel^* = \bm{b} \cdot \bm{B}_\parallel^*$, and $\bm{B}_\parallel^* = \bm{B} +(Bv_\parallel/\Omega_s)\nabla \times \bm{b}$. We use $\bm{B}_\parallel^* \simeq \bm{B}$. The phase-space advection velocities $\dot{\bm{R}} = \{\bm{R},H\}$ and $\dot{v}_\parallel = \{v_\parallel,H\}$ are defined in terms of the Poisson bracket
\begin{eqnarray}
\{F,G\} &=& \frac{\bm{B}^*}{m_s B_\parallel^*} \cdot \left( \nabla F \frac{\partial G}{\partial v_\parallel} - \frac{\partial F}{\partial v_\parallel} \nabla G \right) \nonumber \\* 
&-& \frac{1}{q_s B_\parallel^*} \bm{b} \cdot \nabla F \times \nabla G.
\label{eq:pb-eqn}
\end{eqnarray}
The gyrocenter Hamiltonian is 
\begin{equation}
H_s = \frac{1}{2}mv_\parallel^2 + \mu B + q_s\langle \phi \rangle_\alpha,
\end{equation}
where $\langle \rangle_\alpha$ is the gyro-average. In the long wavelength limit, $\langle \phi \rangle_\alpha = \phi$. In the continuous-time limit, the DG advection scheme we use is energy-conserving. For more details of our computational scheme, see Refs. \citenum{shi2017gyrokinetic,Shi2017thesis,shi2019full}.
\\
\indent We solve for the electrostatic potential using the long-wavelength, gyrokinetic Poisson equation with a linear ion polarization density
\begin{equation}
 -\nabla \cdot \left( \frac{n_{i0}^g q_i^2 \rho_{\mathrm{s}0}^2}{T_{e0}} \nabla_\perp \phi \right) = \sigma_g
 = q_i n_i^g(\bm{R},t) - e n_e(\bm{R},t),
 \label{eq:poisson}
 \end{equation}
where $\rho_{\mathrm{s}0} = c_{\mathrm{s}0}/\Omega_i$ is the ion sound gyroradius and $c_{\mathrm{s}0} = \sqrt{T_{e0}/m_i}$ is the ion sound speed. (Note that the left hand side is actually independent of electron temperature since $\rho_{s0}^2 \propto T_{e0}$, but it is convenient to recover conventional normalizations so that $\nabla_\perp^2$ is normalized by $\rho_{s0}^2$, and $e \phi$ is normalized by $T_{e0}$.) In these simulations, we assume $q_i=e$, which appears to be a reasonable approximation since spectroscopic measurements indicate no ionzation states greater than $Z=1$ for an argon plasma in the Helimak. The background ion guiding-center density $n_{i0}^g$ is taken to be constant in space and time. Nonlinear polarization with a spatially- and time-dependent density is important future work.\\
\indent We artificially increased the electron mass so that $m_i/m_e = \nobreak 400$, which is about 180 times smaller than the actual argon ion-to-electron mass ratio but twice the ratio used in previous fluid simulations.\cite{li2011turbulence} The electron mean free path, $\lambda_{ee} = v_{t,e}/\nu_{ee}$, is the same as experiment since the thermal speed and collision frequency contain the same mass dependency. The increased electron mass does significantly reduce electron parallel thermal conduction, but it is still fast enough that electron temperature is uniform along field lines. \\
\indent We use a non-orthogonal field-line-following coordinate system\cite{beer1995field,hammett1993developments,scott1998global} to model the helical field lines: $z$ is the distance along the field, $x$ is the radial coordinate, and $y$ is the bi-normal coordinate. Field-line curvature enters through the second term in $\bm{B}^*$, which contributes terms proportional to $\nabla \times \bm{b} \cdot \nabla y = -1/x $. The field strength is dominated by the vacuum toroidal field; therefore, we assume $B(x) = B_0 (R_0/x)$. Using the convention for field-line-following coordinate systems\cite{beer1995field}, the direction of increasing $z$ points in the clockwise direction when viewed from above. The Helimak experimental magnetic field points in the counterclockwise, or $-z$, direction. Transformation to a cylindrical coordinate system $(R,\varphi,Z)$ is given by
\begin{eqnarray}
    \label{eq:flux-map1}
    Z &=& -2.0z/L_z + 1.0  \\ 
    \label{eq:flux-map2}
    \varphi &=& \frac{2\pi}{L_y}\left(y + \frac{2.0z}{L_z} - 1.0\right),
\end{eqnarray}
where $L_y$ is the extent of the domain in $y$ and $L_z$ is both the magnetic field line connection length and extent of the domain in $z$. See Fig.~\ref{fig:flux-mapping} for more details. The unit tangent vectors in the figure are defined as $\hat{\bf{e}}_y = \partial \bm{R}/\partial y/|\partial \bm{R}/\partial y|$ and $\hat{\bf{e}}_z = \partial \bm{R}/\partial z/|\partial \bm{R}/\partial z|$.\\
\indent Conducting-sheath boundary conditions\cite{shi2017gyrokinetic,Shi2017thesis,shi2019full} are used in the $z$ direction, where field-lines intersect conducting plates, for the distribution function $f$. These permit local parallel current fluctuations into and out of the wall. Since we assume quasineutrality in our model, we do not resolve the Debye sheath, which contains $n_i > n_e$ and is on the order of a few Debye lengths. The wall is assumed to be just outside the simulation domain. We solve for the sheath potential $\phi_{\mathrm{sh}}$ at the boundary using the gyrokinetic Poisson equation (\ref{eq:poisson}). This sets a cut-off parallel velocity for electrons, $\frac{1}{2}m_e v_c^2 = e \phi_{\mathrm{sh}}$. Electrons with velocities greater than $v_c$ leave the domain, while those with smaller velocities are reflected. Dirichlet boundary conditions are used in $x$ for the potential ($\phi = 0$), and periodic boundary conditions are used in $y$ for $f$ and $\phi$. 
\begin{table}[h]
\caption{\label{tab:table1}Summary of Helimak simulation parameters}
\begin{ruledtabular}
\begin{tabular}{ll}
Background magnetic field & $B_0 =  0.1$ T \\
Background density & $n_0 = 10^{16}$ m$^{-3}$\\
Background electron temperature & $T_{e0} = 10$ eV\\
Background ion temperature & $T_{i0} = 1$ eV\\
Density gradient scale length & $L_n = 0.1$ m \\
Electron gyroradius & $\rho_e = 1.02 \times 10^{-2}$ m \\
Electron mean free path & $\lambda_{ee}$ = 98.4 m \\
Ion gyroradius & $\rho_i = 6.46 \times 10^{-3}$ m \\
Ion sound gyroradius & $\rho_{\mathrm{s}0} = 2.04 \times 10^{-2}$ m \\
Ion sound speed & $c_\mathrm{s} = 4.90 \times 10^{3}$ m/s \\
Ion mean free path & $\lambda_{ii}$ = 1.39 m \\
Ion cyclotron frequency & $\Omega_{ci} = 2.40 \times 10^5$ s$^{-1}$ \\
Ion-ion collision frequency &$\nu_{ii} = 1.11 \times 10^3$ s$^{-1}$ \\
Ion-neutral collison frequency &$\nu_{i0}\ge 185$ s$^{-1}$
\end{tabular}
\end{ruledtabular}
\end{table}

\begin{figure}
    \centering
    \includegraphics[width=0.5\textwidth]{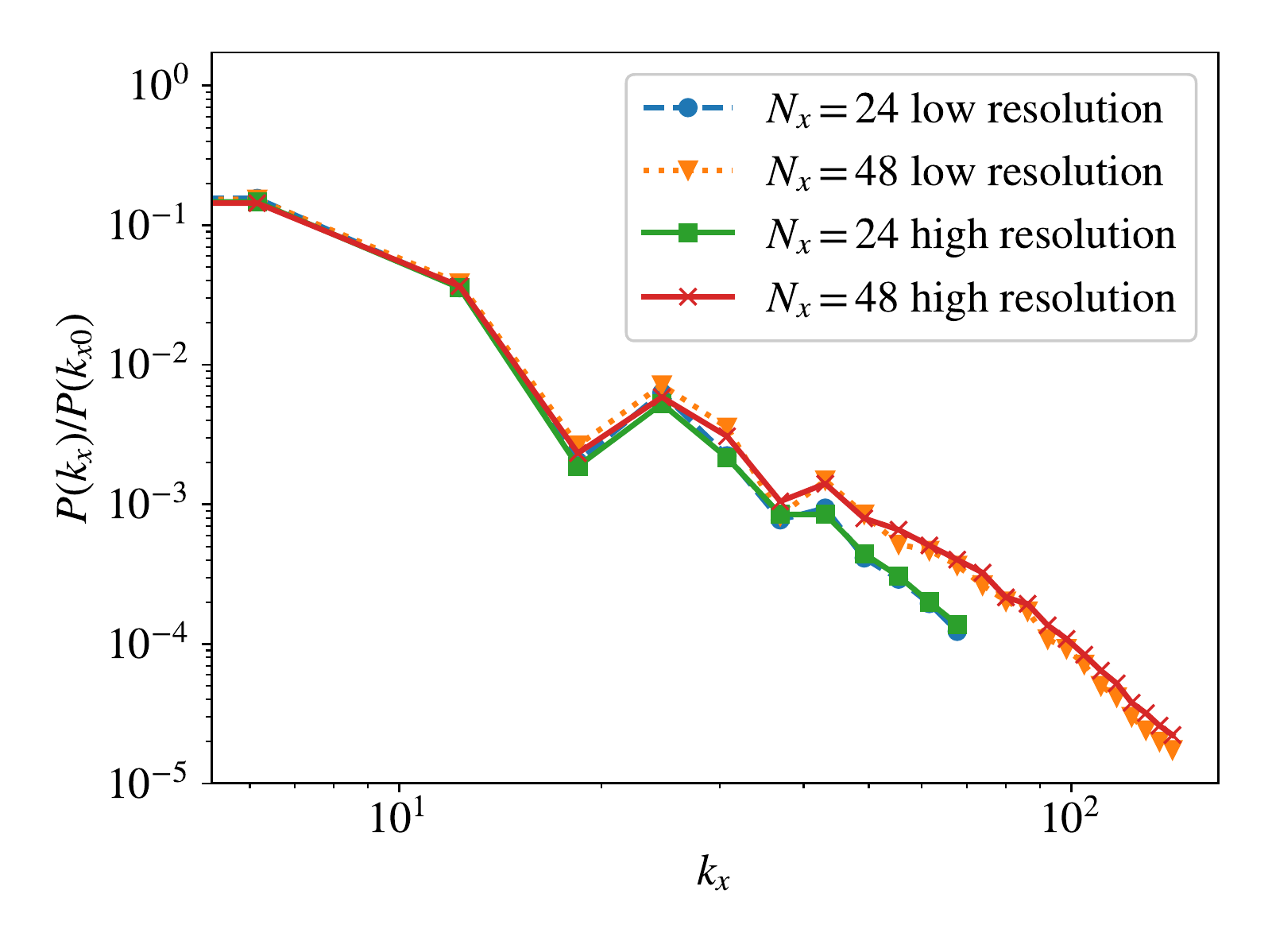}
    \caption{Power in the Fourier transform in $x$ of electron density fluctuations is compared for various resolutions. The power has been averaged in $y$, $z$, and in time from 10 to 16 ms. The low resolution case has $(N_x,N_y,N_z,N_{v_\parallel},N_\mu) = (N_x,12,8,10,5)$ and the higher resolution case has $(N_x,N_y,N_z,N_{v_\parallel},N_\mu) = (N_x,24,16,10,5)$. The decay of the power at high $k_x$ indicates that our simulations are not changed significantly with increased resolution.}
    \label{fig:kx-converge}
\end{figure}

The Helimak simulations with Gkeyll evolve the electron and ion distribution functions for an argon discharge. We chose simulation parameters based on typical experimental parameters\cite{gentle2008texas,gentle2010comparison} and previous fluid simulations,\cite{li2009plasma,li2011turbulence} using $B_0 = 0.1$ T and $n_{i0}^g = n_{e0} = 10^{16}$~m$^{-3}$. Since the electron--ion thermal equilibration time is much longer than a charge-exchange time and particle confinement times for parallel loss,  $T_{i0} \ll  T_{e0}$ in the experiment. We use $T_{e0} = 10$ eV based on experimental measurements. Previous work\cite{perez2006drift,li2009plasma} estimates $T_{i0} \sim 0.1$~eV, though this has not been accurately measured. 
Argon that is ionized at different locations will be accelerated for different lengths of time towards the device ends by the pre-sheath potential, which varies along a field line. This will produce an ion temperature that is a significant fraction of the pre-sheath potential drop, which is of order $T_{e0}$. Therefore, we use $T_{i0} = 1$ eV. The equilibrium temperatures define the velocity grid and appear in the the source terms, which we explain in more detail below. The particle temperatures $T_i$ and $T_e$ vary spatially and are evolved self-consistently in time. A gyrokinetic model is justified for these parameters since the ion-ion collision frequency, $\nu_{ii} = \nobreak 1.112 \times \nobreak 10^3$~s$^{-1}$, is much less than the ion cyclotron frequency, $\Omega_{ci} = \nobreak 2.40 \times 10^5$ s$^{-1}$, and the ion gyroradius, $\rho_i = \nobreak 6.46 \times \nobreak 10^{-3}$ m, is much less than the density gradient scale length, $L_n = 0.1$ m. See Table \ref{tab:table1} for a summary of key simulation parameters. \\
\indent The toroidal vacuum chamber of the Helimak is rectangular with a radial extent 0.6 m $\leq R \leq 1.6$ m and a vertical height $H = 2.0$ m. By changing the current through the vertical field coils, the magnetic-field-line connection length $L_z$ can be varied from 12 m to 500 m. In our simulations, we use a slab-like metric with a spatial Jacobian that is independent of $x$. To simulate the entire volume of the Helimak, the configuration space extents are $x \in [0.6,1.6]$ m and $z \in [-L_z/2,L_z/2]$. $L_y$ varies inversely with the connection length: $L_y = H/N$, where $N = L_z / (2 \pi R_0) $ is the number of field-line turns and $R_0 = 1.1$ m. This gives $y \in [-L_y/2, L_y/2]$. In the experiment, there is magnetic shear due to the variation in the field with respect to $R$, which is not included in this model. This also means that the connection length $L_z$ and binormal length of periodicity $L_y$ do not vary in $x$ as they realistically should. The parallel velocity-space extents are given by $v_{\parallel,e} \in [-v_{e,\text{max}}, v_{e,\text{max}}]$ and $v_{\parallel,i} \in [-v_{i,\text{max}}, v_{i,\text{max}}]$, where $v_{e,\text{max}} = 4 v_{te} = 4 \sqrt{T_{e0}/m_e}$ and $v_{i,\text{max}}= 6 c_\mathrm{s} = 6 \sqrt{T_{e0}/m_i}$. The perpendicular velocity-space extents for each species are given by $\mu_s \in [0,3 m_s v_{s,\text{max}}^2/(16 B_0)]$. The grid resolution is $(N_x,N_y,N_z,N_{v_\parallel},N_\mu) = (48,24,16,10,5)$ with uniform spacing. The numerical solutions for the distribution function and the Hamiltonian are projected onto piecewise-linear basis functions using the discontinuous Galerkin scheme. Because the gyrokinetic system is high in number of dimensions and, hence, computationally intensive, convergence tests were performed only in the $x$ direction for a low-resolution, ($N_x$,$N_y$,$N_z$,$N_{v_\parallel}$,$N_\mu$) = ($N_x$, 12, 8, 10, 5), and a higher-resolution, ($N_x$,$N_y$,$N_z$,$N_{v_\parallel}$,$N_\mu$) = ($N_x$, 24, 16, 10, 5), case. Unpublished resolution studies in other directions have been done in the past for various conditions. Power in the Fourier transform in $x$ of electron density fluctuations are compared for different resolutions in Fig.~\ref{fig:kx-converge}. Fluctuations are calculated as $\tilde{n} = n - \langle n \rangle_y$, where the last term denotes an average in the $y$-direction. The power has been averaged in $y$, $z$, and in time from 10 to 16 ms. The decay of the power at high $k_x$ indicates that the simulations are not changed significantly with increased resolution. Furthermore, equilibrium profiles and turbulence statistics were similar for the different cases. \\
\begin{figure*}
    \includegraphics[width=\textwidth]{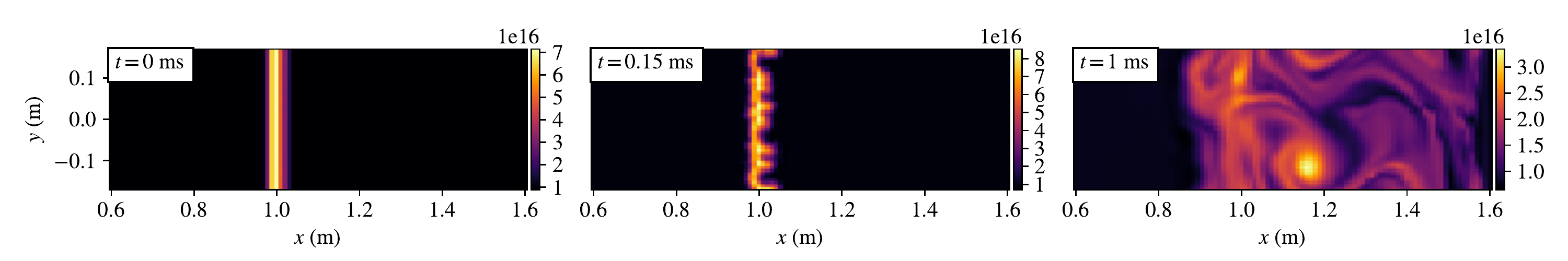}
    \caption{Snapshots of ion guiding-center density (m$^{-3}$) in the $(x,y)$ plane at the midpoint in $z$ of the field-line-following coordinate system. Initial conditions are depicted at 0 ms (left), turbulent structures are visible at 0.15 ms (center), and a non-linear turbulent state is apparent at 1 ms (right).}
    \label{fig:3n}  
\end{figure*}
\begin{figure}
    \centering
    \includegraphics[width=0.5\textwidth]{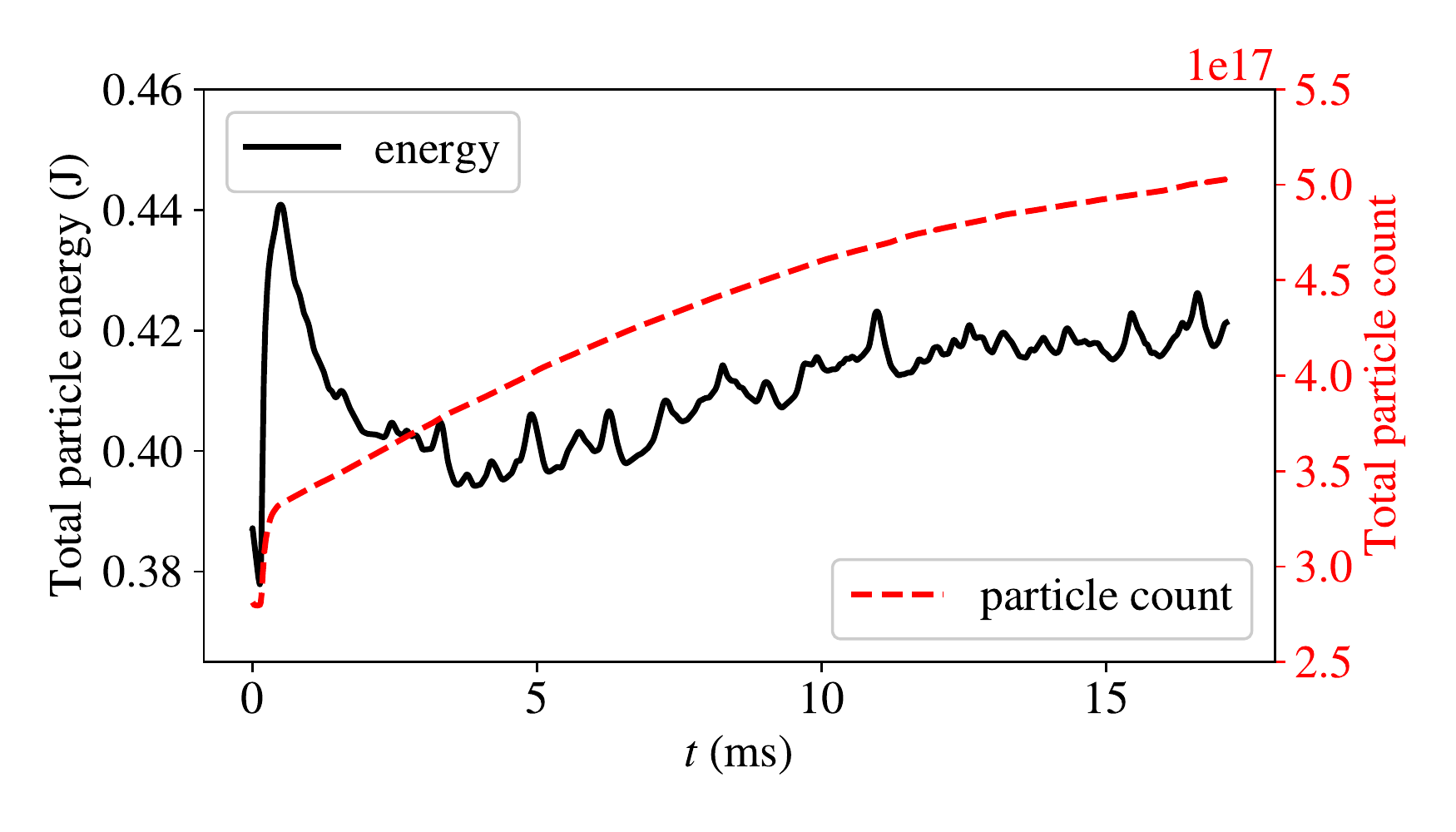}
    \caption{Total particle count and the total particle energy in the simulation. Particle energy for a single species is calculated as the integration of the product of the particle Hamiltonian and distribution function over all phase space. Time-averaged profiles and turbulence statistics were calculated from 10 to 16 ms.}
    \label{fig:sat}
\end{figure}
\begin{figure*}
    \includegraphics[width=\textwidth]{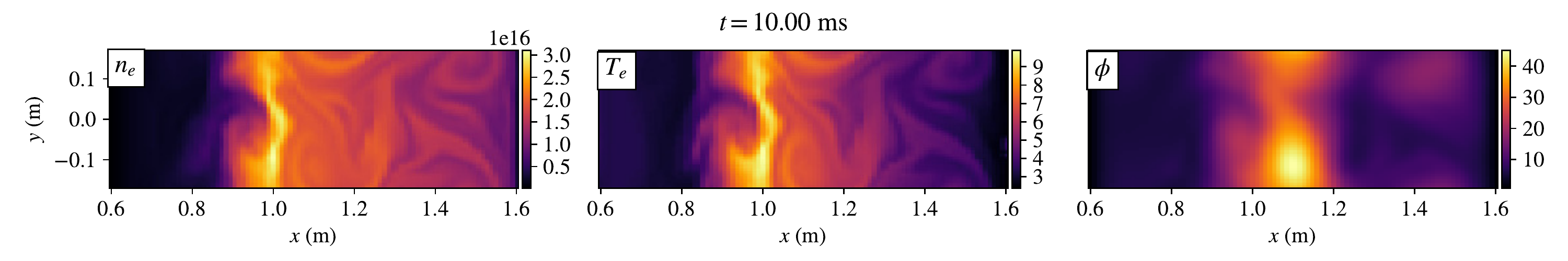}
    \caption{From left to right, electron density (m$^{-3}$), electron temperature (eV), and electrostatic potential (V) in ($x,y$) in the field-line-following coordinate system at the midpoint in $z$. (Multimedia view)}
    \label{fig:nTphi}
\end{figure*}
\indent In the Helimak, electrons are preferentially heated by electron-cyclotron and upper-hybrid resonances. The latter is believed to dominate and is localized from 1.0--1.1~m radially.\cite{Williams2017thesis} The upper-hybrid resonance heating is difficult to model accurately due to its density dependence. The source is also challenging to model because the particle source rate and power deposition into the plasma are not accurately known. It is estimated that over 90$\%$ of the 6 kW input power is lost to radiation by electron-impact excitation of neutral argon. This process depends on the neutral density profile. We estimated the total radiative cooling using the predicted radiative cooling rate $L_Z$ in Ref. \citenum{mavrin2017radiative}. The total power radiated in the device is given by $P_{\text{rad}}\mathcal{V}= L_Z n_e n_{\text{imp}} \mathcal{V} \simeq 5.5$ kW, assuming $T_e = 10$ eV, $n_e=1 \times 10^{16}$ m$^{-3}$, and the total Argon density, including neutrals, is $n_{\text{imp}}=4 \times 10^{17}$ m$^{-3}$. $\mathcal{V}$ is the total volume of Helimak vacuum vessel. Neutral interactions and radiation are neglected in our present source model except indirectly through the usage of a net source rate, as described below. We estimate the ion-neutral collision frequency $\nu_{i0} \ge 185$ s$^{-1}$ from the electron-neutral frequency\cite{gentle2014turbulence}, which differs by a factor of $\sqrt{m_e/m_i}$, assuming ions travel at the sound speed along field lines. This is much less than the gyrofrequency, though close to the ion transit frequency $L_z/(2c_\mathrm{s}) = 245$ s$^{-1}$. Thus, the effect of neutrals is important and a feature we plan to include in future work.\\ 
\indent We currently employ the simplified source model used in Refs.~\citenum{shi2017gyrokinetic,Shi2017thesis,shi2019full}. For both species, we assume a source of the form
\begin{eqnarray}
    \label{eq:source-profile}
    S_s(\bm{R}, v_\parallel, \mu) = S_0 \exp[-(x - x_\mathrm{src})^2 / (2\sigma_\mathrm{src}^2)] \nonumber \\ 
\times F_M(v_\parallel, \mu; T_{s,\mathrm{src}}),
\end{eqnarray}
where $s$ refers to species, $F_M$ is a normalized, non-drifting Maxwellian,  $x_\mathrm{src} = 1.0$ m is the source location, and $\sigma_n = 0.01$ m is the source width. $T_{s, \mathrm{src}}$ is the source temperature. The source is independent of $y$, $z$, and $t$. Previous fluid simulations use $\sigma_\mathrm{src} \sim 0.1$ m. We chose a narrower source because the RF resonance is probably very narrow, though the location of the upper-hybrid resonance fluctuates as the density fluctuates. Equilibrium profiles are broadened by turbulence and wider than the assumed source width. Low-resolution scans in $\sigma_\mathrm{src}$ from 0.1 to 0.01 m produced little effect on resulting equilibrium profiles. \\
\indent Since the actual particle and power source rates are not well known, we estimate the particle source rate $S_0$ and source temperature $T_{s,\mathrm{src}}$ using a 1D fluid transport model (see Appendix B in Ref.~\citenum{Shi2017thesis} or Appendix A in Ref.~\citenum{shi2019full}) to approximately produce the observed quasi-steady-state density and temperature profiles. We assume a balance of source rates and losses using
\begin{equation}
\frac{d\mathcal{N}}{dt} = \int d\bm{R} \, \left[ S_n(x) - \frac{n(x,z)}{\tau_\parallel} \right]= 0,
\label{eq:steadystate}
\end{equation}
where $\mathcal{N}$ is the total number of particles and $\tau_\parallel =\nobreak L_z /(2 c_\mathrm{s})$ is the parallel transit time. We assume a steady-state density profile with a variation in $z$ based on a 1D steady-state single fluid calculation:  
\begin{eqnarray}
\label{eq:dens-source}
n(x,z) = n_{p} \exp \left [-(x - x_\mathrm{src})^2 / \left (2\sigma_n^2 \right ) \right ] \nonumber \\ 
\times \frac{1 +\sqrt{1 - z^2/(L_z/2)^2}}{2}. 
\end{eqnarray} 
The peak density $n_p$ is calculated such that $\int d\bm{R} \,\, n(x,z) / \mathcal{V}$ is equal to the background density. We define the density source for each species as $S_{s,n} (x) \equiv \int d\mathbf{v} \, S_s(\bm{R}, \mu, v_\parallel)$ and use Eqs.~\ref{eq:steadystate} and \ref{eq:dens-source} to  estimate $S_0 \approx 9.77 \times 10^{19}$ m$^{-3}$s$^{-1}$ for the source amplitude. We estimate  $T_{s,\mathrm{src}}=\frac{5}{3}T_{s0}$ for the source temperature in Eq.~\ref{eq:source-profile} based on the temperature at the midpoint in $z$ of a 1D steady-state fluid calculation.\cite{Shi2017thesis,shi2019full} However, that model assumes only convective outflows and neglects the fast parallel thermal transport of electrons, so it was necessary to double this value for electrons ($T_{e,\mathrm{src}}=\frac{10}{3}T_{e0})$ to approximately match experimental electron temperature profiles. \\
\indent We expect the quasi-steady-state conditions of the simulation to be independent of initial conditions. Therefore, we chose initial conditions with a narrow density gradient scale length and a non-zero flow velocity as a function of the parallel coordinate, computed from simplified 1D fluid models following Ref. \citenum{Shi2017thesis} and \citenum{shi2019full}. The density initial conditions are pictured in the left plot of Fig.~\ref{fig:3n}. We also used a density floor of 10$\%$ to minimize unphysical negative values in our full-$f$ distribution function. The code includes a scheme to correct for negative values of the distribution function, detailed in Refs. \citenum{shi2017gyrokinetic} and \citenum{Shi2017thesis}, which adds a small amount of numerical heating ($\sim 10\%$) to the simulations in the quasi-steady state. A new version of the code is under development and has a more robust method of preventing negative values of the distribution function.\\ 
\indent Scans of the ion-to-electron mass ratio were performed in previous helical open-field-line simulations with Gkeyll and resulted in no significant changes to turbulence statistics.\cite{Shi2017thesis,shi2019full}  The difference between the true and reduced mass ratio for argon is much larger in the case of the Helimak. The large parallel ion transit time requires longer simulation times to reach saturation and is computationally expensive. By solving for linear perturbations in the parallel current, $\tilde{j}_\parallel$, near the sheath and using $j_{\parallel e} + j_{\parallel i} = 0$, we estimated that the reduced ion-to-electron mass ratio would under-predict the response of electrostatic potential to temperature fluctuations by approximately $50\%$. A comparison of  low-resolution simulations, $(N_x,N_y,N_z,N_{v_\parallel},N_\mu) = (24,12,8,10,5)$, with the true and reduced mass ratio supports this prediction. A new modal discontinuous Galerkin version of Gkeyll with automatic code generation is under development and is much faster. It should allow routine use of higher ion-to-electron mass ratios in the future.

\section{Simulation results} \label{sec:results}
\indent We ran the simulation using the Skylake (SKX) compute nodes on the Stampede2 cluster at the Texas Advanced Computing Center. A simulation of a 16 ms argon discharge for $L_z = 40$ m required approximately 180,000 CPU-hours. This is about 4 times the ion transit time, $\tau_\parallel = L_z/(2c_\mathrm{s}) = 4.084$ ms, which is the time it takes an ion traveling at its characteristic sound speed, with $T_{e0} = 10$ eV, to traverse half the magnetic-field-line connection length. We compared with experimental data from a discharge with a similar connection length at $R_0 = 1.1$ m, corresponding to $B_Z \simeq 0.005$ T and $B_Z/B_\varphi \simeq 0.05$. We selected this configuration as it results in the shortest connection length for which magnetic field lines terminate on conducting plates on the bottom and top of the vacuum chamber. \\
\indent Figure \ref{fig:sat} shows the total particle count and total energy, $\mathcal{E}_\mathrm{tot} = \mathcal{E}_e + \mathcal{E}_i$, of the simulation. Particle energy for a single species is calculated as the integration of the product of the particle Hamiltonian and distribution function over all phase space:
\begin{equation}
    \mathcal{E}_s = \int d\mathbf{z} \, H_s \, f_s(\mathbf{z}).
    \label{eq:energy}
\end{equation}
Particle count is still increasing, though beginning to level-off by the end of the simulation. The energy appears to saturate around 10 ms. The large spike in energy at early time is due to numerical heating added from the scheme to correct for negative distribution function values. The nearly constant energy but increasing particle count indicates a slight decrease in temperature. We calculate equilibrium profiles and turbulence statistics over the interval from 10 to 16 ms.\\
\begin{figure}
    \includegraphics[width=0.5\textwidth]{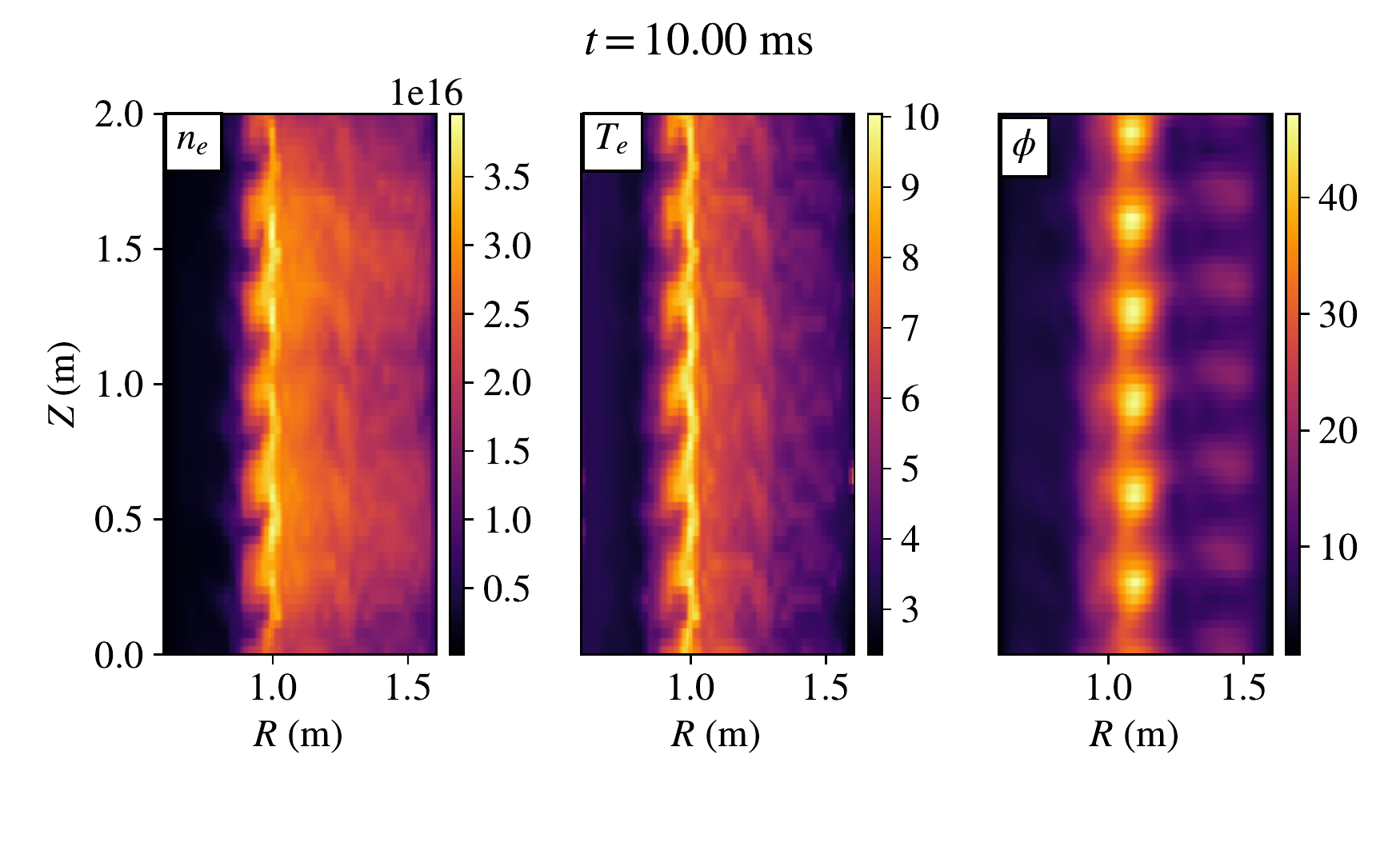}
    \caption{From left to right, electron density (m$^{-3}$), electron temperature (eV), and plasma potential (V) in the lab coordinates $(R,Z)$ at a slice in the toroidal coordinate $\varphi$. (Multimedia view)}
    \label{fig:nTphi-lab}
\end{figure}
\indent In the simulation, turbulent structures are visible at 0.15 ms, as seen in the center plot of figure \ref{fig:3n}, which contains snapshots of the ion guiding-center density in the non-orthogonal field-line-following coordinate system. The right-hand plot shows a non-linear turbulent state at 1 ms. According to the energy plot in Fig.~\ref{fig:sat}, a quasi-steady state is reached at approximately 10 ms. Figure \ref{fig:nTphi} (Multimedia view) depicts electron density, electron temperature, and electrostatic potential at $t=$ 10 ms in the field-line-following coordinate system at the midpoint along $z$. Using the transformation in Eqs.~\ref{eq:flux-map1} and \ref{eq:flux-map2}, 
this data was mapped to the cylindrical coordinate system $(R,\varphi,Z)$ and is shown in Fig.~\ref{fig:nTphi-lab} in the $(R,Z)$ plane at a slice in $\varphi$. $N=L_z/(2\pi R_0)=5.8$ is the number of toroidal turns of the field line, and this periodicity is apparent in the figure. For more details of the mapping, see Fig.~\ref{fig:flux-mapping}. \\
\begin{figure}
    \includegraphics[width=0.5\textwidth]{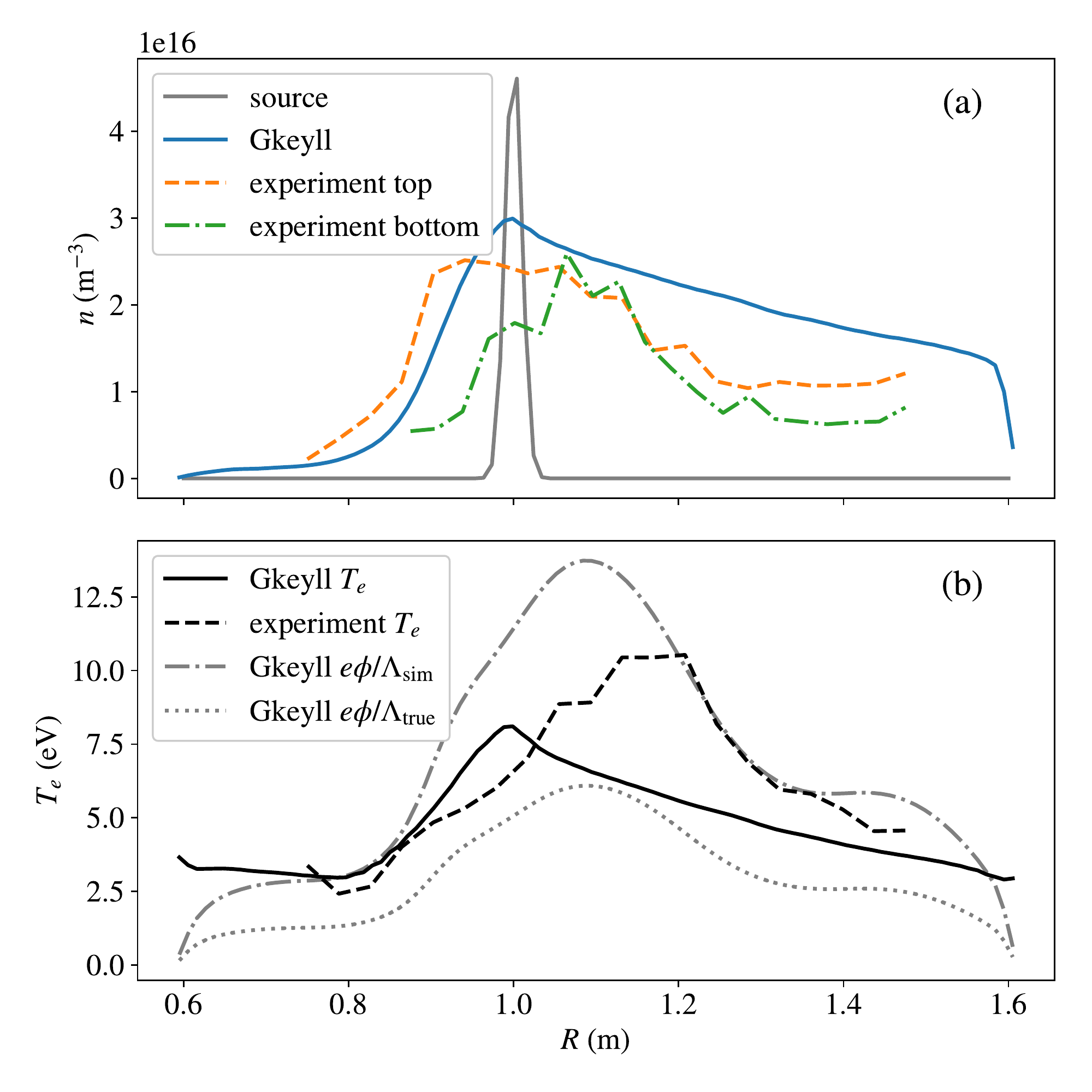}
    \caption{Comparison of (a) the electron density profiles shown with the narrow source and (b) electron temperature profiles shown with the simulated plasma potential, assuming an adiabatic electron response. $\Lambda=\text{ln}\sqrt{m_i/(2\pi m_e})$, where $\Lambda_{\text{true}}$ uses the real ion-to-electron mass ratio and $\Lambda_{\text{sim}}$ uses the reduced ion-to-electron mass ratio of 400. $e \phi / \Lambda_{\rm sim}$ is not exactly equal to $T_e$ because the conducting sheath boundary condition allows currents to flow in and out of the walls.}
    \label{fig:eqprof}
\end{figure}
\indent Figure \ref{fig:eqprof} compares equilibrium profiles of simulation and experimental data. Experimental Langmuir probes are located on conducting plates at the top and bottom of the vacuum vessel. The probe measurements of the equilibrium profiles have about a 50\% uncertainty, and experimental data is generally fit to a smooth curve. There is a top--bottom asymmetry in experimental density profiles. This is not present in the simulation, since our model is symmetric in $z$. Thus, in all comparisons we calculate simulation data at the minimum of the $z$ domain ($z=-L_z/2$). One reason for this asymmetry could be the vertical, $Z$ in the lab frame or $z$ in the simulation, \ExB\ drift, which we presently neglect in our model. Discussing this issue is complicated because of the non-orthogonal coordinate system that is used, but to be precise, the advection due to a radial electric field, $V_{E \times B} \cdot \nabla 
= V_{E \times B} \cdot \nabla y \, \partial / \partial y 
+ V_{E \times B} \cdot \nabla z \, \partial / \partial z$
is approximated by 
$V_{E \times B} \cdot \nabla y \partial / \partial y$, which advects in $y$ at fixed $z$ and, thus, at fixed height $Z$. This is a common assumption for core flux-tube codes assuming $k_{\perp} \gg k_{||}$, but can break down for long wavelength components or small magnetic pitch angle.
Based on experimental flow data from Ref.~\citenum{gentle2010comparison}, the vertical \ExB\ flow is 2--3 times the vertical component of the sonic outflows along the field lines for simulated parameters. This factor increases as the toroidal field component $B_T$ increases relative to the vertical component $B_Z$ and may influence observed density profiles. Additionally, the microwave waveguide enters the machine at the bottom and may explain the similarity in bottom density profiles for different pitch angles which is not observed for top profiles.\cite{Williams2017thesis} Experimental temperature profiles are fairly top--bottom symmetric and centered around 1.1 m, irrespective of pitch angle.\cite{Williams2017thesis} Since electron parallel heat conduction is rapid, the electron-temperature steady-state profiles would be less affected by the vertical \ExB\ drift or the location of the microwave guide. The vertical component of \ExB\ flow is likely important in transport balances, though not as important for turbulence fluctuations. A full analysis of the effects of mean vertical \ExB\ flows on the asymmetry in the experimental profiles involves solving a complex non-linear equation and is not within the scope of this paper. \\
\indent We model the source for each species using a single Maxwellian with a flat temperature profile and a density profile centered at 1.0 m. We chose the location of the source based on experimental steady-state density profile. Figure \ref{fig:eqprof}(a) depicts the narrow electron density source and resulting equilibrium profiles which were broadened by turbulence. The simulation density peak is close in magnitude to experimental profile peaks. However, the simulation fails to capture the steep gradient just to the right of the peak and the flat profile at higher radii that is present in the experiment. This might be improved by implementing a density-weighted Poisson equation, instead of the current one Eq.~\ref{eq:poisson}, in which the background ion guiding center density $n_{i0}^g$ is constant in space and time. This approximation is similar to the Boussinesq approximation used in many fluid models.\cite{li2009plasma,li2011turbulence,li2017edge} Some simulations have relaxed this approximation\cite{francisquez2017global,dudson2016verification,halpern2016gbs} but have not been tested with Helimak parameters.

Figure \ref{fig:eqprof}(b) shows that the simulation produced a temperature profile centered to the left of the experimental profile. The peak magnitude and the gradient to the right of the peak were both less than experiment. The simulated plasma potential profile is shown in the same plot as $e\phi/\Lambda$, where $\Lambda=\text{ln}\sqrt{m_i/(2\pi m_e})$ for the reduced ion-to-electron mass ratio and the true mass ratio. An estimate of the simulated floating potential using $\phi_{\text{fl}} \simeq \phi - T_e\Lambda/e$ agrees qualitatively with experimental measurements of the same quantity.\cite{gentle2008texas,gentle2010comparison,Williams2017thesis} \\
\begin{figure}
    \includegraphics[width=0.5\textwidth]{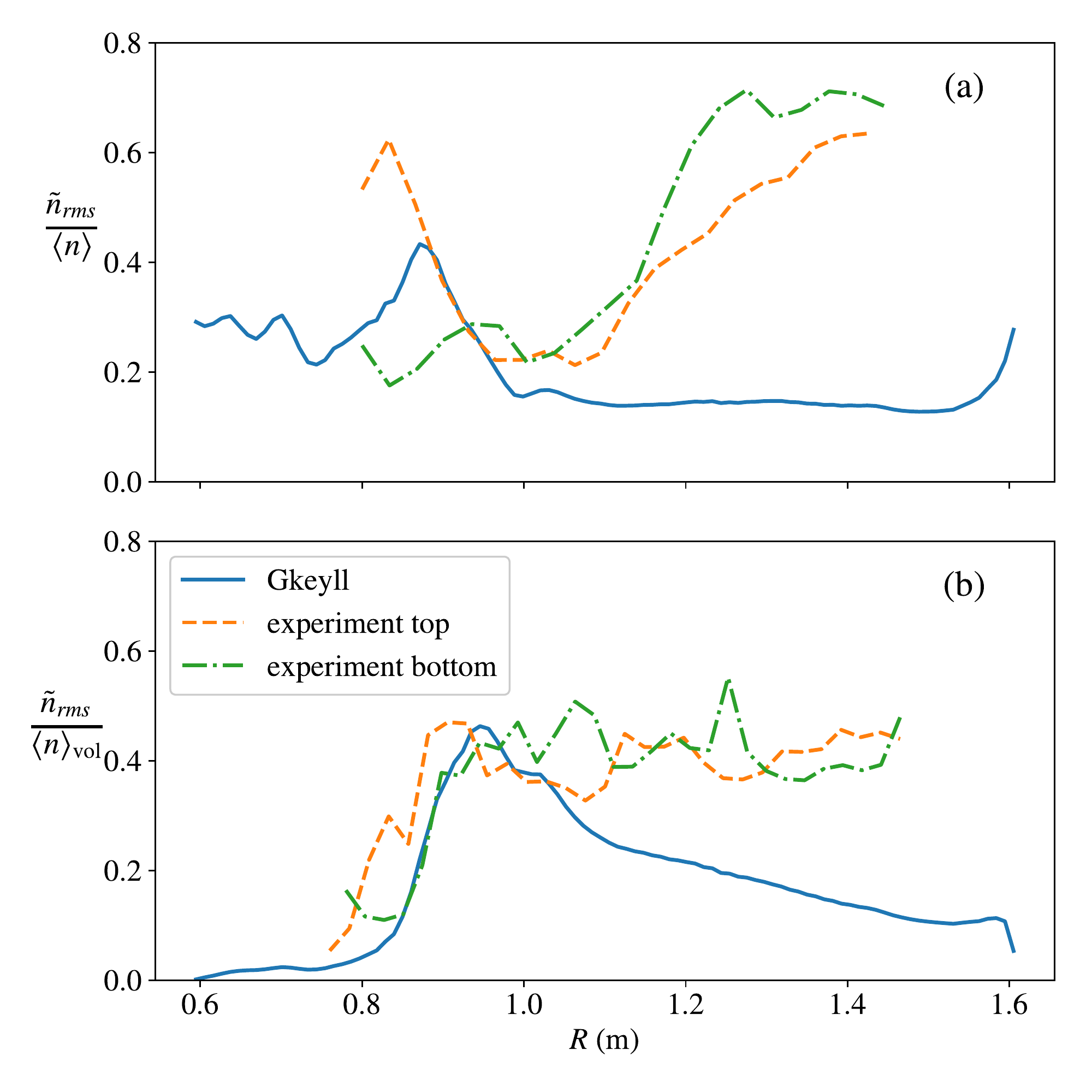}
    \caption{Electron density fluctuation profiles normalized to (a) the local mean density $\langle n \rangle(R)$, which is averaged in time and in the azimuthal angle and (b) normalized to volume-averaged density $\langle n \rangle_{\mathrm{vol}}$, which is averaged in time and over the entire volume.}  
    \label{fig:dn-profile}
\end{figure}
\indent The simulated turbulence profiles come closer to experimentally measured levels than previous fluid simulations.\cite{li2009plasma} Turbulence levels are calculated as the root-mean-square of density fluctuations, $\tilde{n}=n - \langle n \rangle$, normalized to the mean electron density $\langle n \rangle$. In the Helimak, ion saturation current measurements, $I_{\rm{sat}} = n_e \sqrt{T_e}$, are used as a proxy for density in density fluctuation statistics and have very small associated errors. For comparison, we also calculate simulation density fluctuations using $I_{\rm{sat}} = n_e \sqrt{T_e}$ but, henceforth, refer to all density fluctuations as $\tilde{n}$, as opposed to $\tilde{I}_{\rm{sat}}$, as is typically done in Helimak experimental papers. In the simulation, angle brackets $\langle \rangle$ denote an average in time, performed from 10 to 16 ms, and an average in the azimuthal angle. Plot \ref{fig:dn-profile}(a) is normalized to a local mean density, while \ref{fig:dn-profile}(b) is normalized to a global mean density. The latter shows slightly better agreement, though both differ at large radii. Using a density-weighted Poisson equation may also improve agreement here.\\
\begin{figure}
    \includegraphics[width=0.50\textwidth]{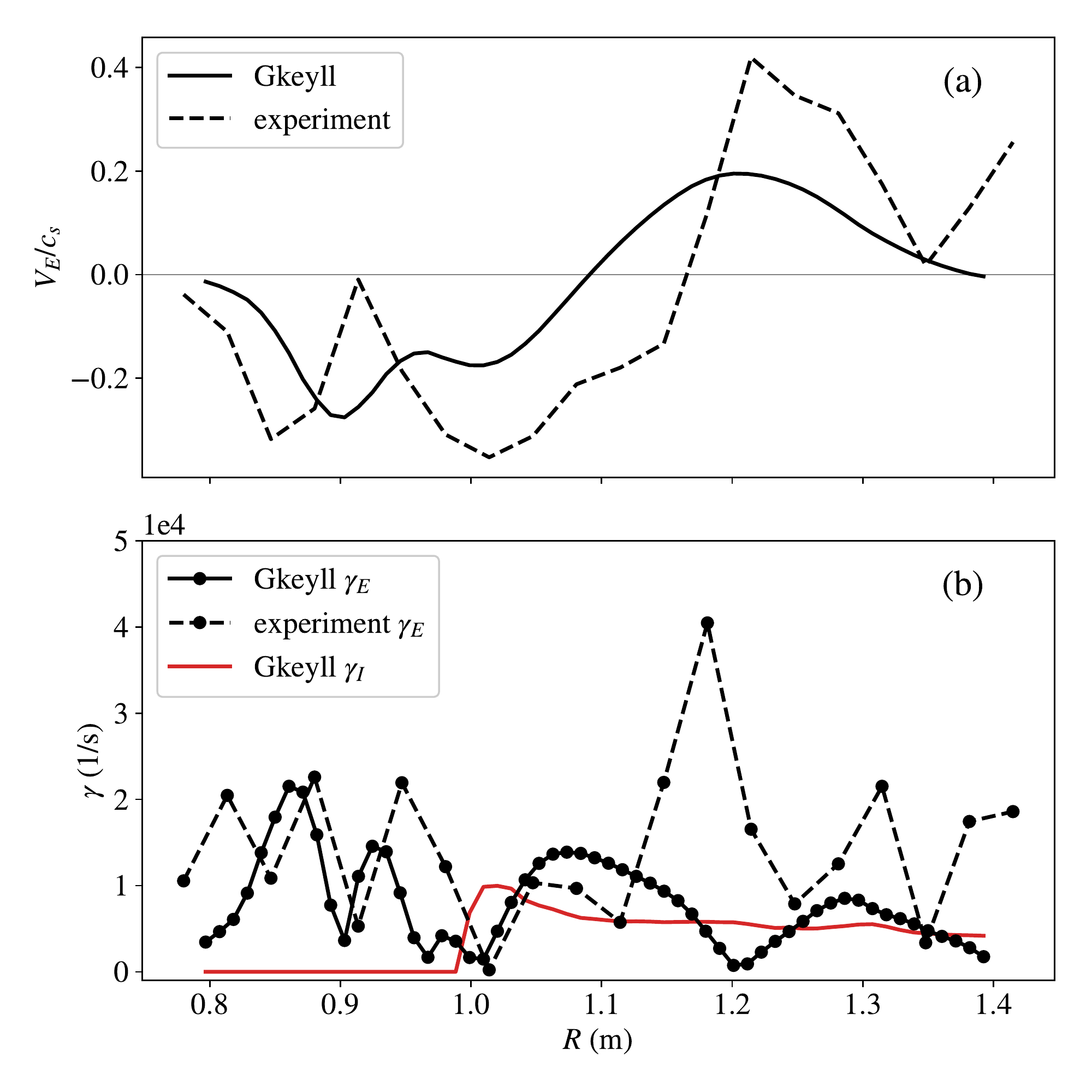}
    \caption{(a) The \ExB\ drift velocity $V_E$ normalized to the ion sound speed and (b) the absolute value of the shear $\gamma_E$ compared to the local interchange growth rate $\gamma_I$. The experimental velocity shear was estimated from the $T_e$ data, assuming $\phi_0 \approx \Lambda T_e/e$.}
    \label{fig:ve-shear}
\end{figure}
\indent The plasma potential equilibrium profile generates a sheared flow perpendicular to the magnetic field. This flow is calculated as $V_E = -(d\phi_0/dx)/B_0$, with the negative sign resulting from the direction of the experimental magnetic field, and the corresponding shear as $\gamma_E = |dV_E/ dx|$. As mentioned above, we currently neglect the vertical \ExB\ in our simulation, and the flow in \ref{fig:ve-shear}(a) is in the $y$, or toroidal, direction. We approximate the experimental flow and shear from the experimental temperature profiles by assuming $\phi_0 \approx \Lambda T_e/e$, though this calculation overestimates the flow velocities as compared with published experimental measurements.\cite{gentle2010comparison,li2011turbulence} We note that this flow is mostly in the vertical direction. Previous research has shown that velocity shear in strongly magnetized plasmas has a stabilizing effect by breaking apart turbulent eddies.\cite{terry2000suppression}
Therefore, we compare shear to the local linear growth rate of the interchange instability $\gamma_I = c_s/\sqrt{R L_p}$ in Fig.~\ref{fig:ve-shear}(b). The pressure gradient scale length is defined as $1/L_p = -(dp/dx)/p$. Turbulence statistics, such as power spectra and auto-correlation functions, are calculated at radial locations where $V_E \simeq 0$, at $R=1.09$ m for the simulation and $R=1.17$ m for the experiment, and where $V_E$ is approximately maximal, at $R=1.20$ m for the simulation and $R=1.25$ m for the experiment. \\
\begin{figure}
    \includegraphics[width=0.45\textwidth]{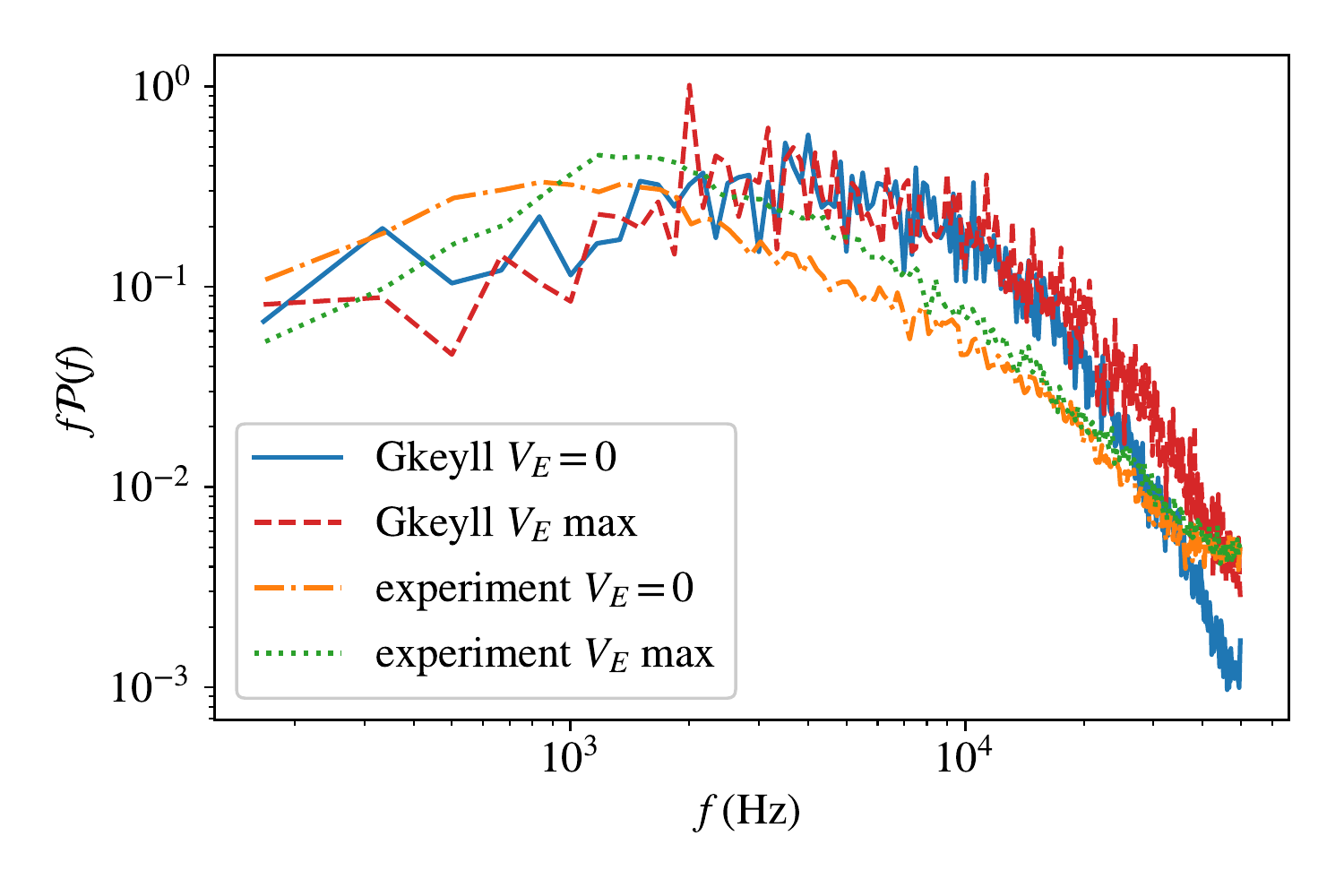}
    \caption{Power spectra of electron density fluctuations, normalized to total energy, for simulation and experimental data at radial locations where $V_E$ is zero and maximal.}
    \label{fig:power-spec}
\end{figure}
\indent The power spectra are calculated from the Fourier transform of the electron density fluctuations $\tilde{n}=n - \langle n \rangle$ with respect to time. Since the density fluctuations are not periodic at the boundaries $t_{\text{min}}$ and $t_{\text{max}}$ of this time series, we multiply the density fluctuations by a sinusoidal window prior to performing the FFT: \\
\begin{equation}
    \tilde{n}_w(t_i) = \left[1 - \text{cos}\left( \frac{2 \pi}{N_t} i \right)\right]\tilde{n}(t_i).
\end{equation}
The integer $i$ runs from 0 to $N_t$, ensuring that the time series is 0 and thus periodic. The power spectra is then normalized to the total energy
\begin{equation}
    \label{eq:power-spec}
    \mathcal{P}(f) = \frac{\langle |\tilde{n}(f)|^2 \rangle}{\sum_f \Delta f |\tilde{n}(f)|^2},
\end{equation}
where $f$ denotes frequency and not the distribution function. The frequency grid spacing $\Delta f$ is 168~Hz. The same window and sampling frequency were used for both simulation and experimental data. The experimental data uses a longer time series and was binned to 6 ms intervals and then averaged. \\
\indent Figure \ref{fig:power-spec} compares the experimental and simulated power spectra multiplied by the frequency, $f\mathcal{P}(f)$, at radial locations where \ExB\ flow is zero and maximal. For the radial values where flow is maximal, the experimental spectrum is weakly peaked near 1.2 kHz, in accordance with previous results,\cite{perez2006drift,li2009plasma} while the simulation is a bit noisier and peaks at a higher frequency, in the 2-4 kHz range. The simulation spectra appear to have less relative power in frequencies just below the peak and greater power above the peak when compared to the respective experimental spectra. Furthermore, simulation spectra decay more quickly at high frequencies.
\begin{figure}
    \includegraphics[width=0.45\textwidth]{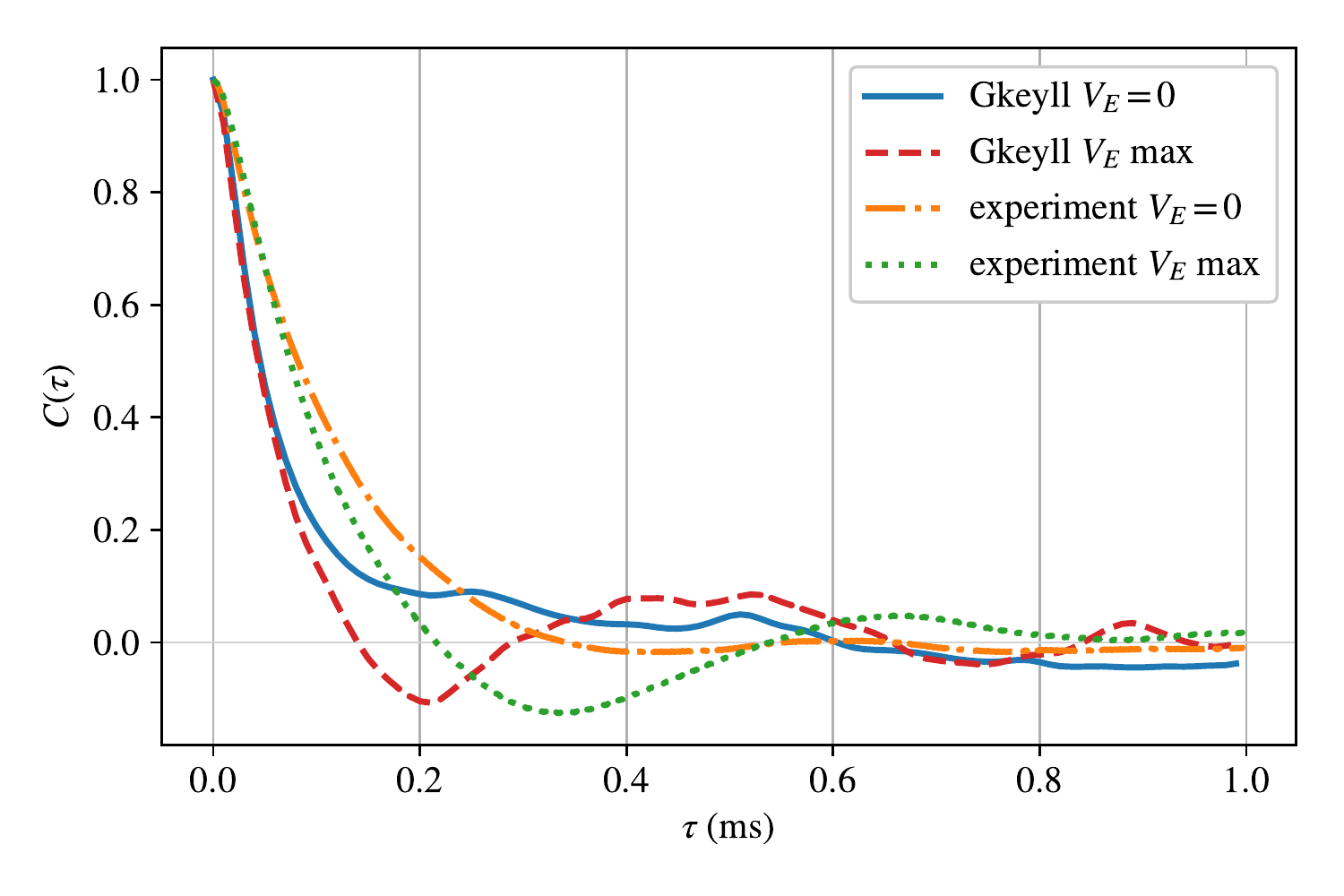}
    \caption{Comparison of autocorrelation functions for simulation and experimental data at radial locations where $V_E$ is zero and maximal. }
    \label{fig:autocorr}
\end{figure}
\begin{figure}
    \includegraphics[width=0.45\textwidth]{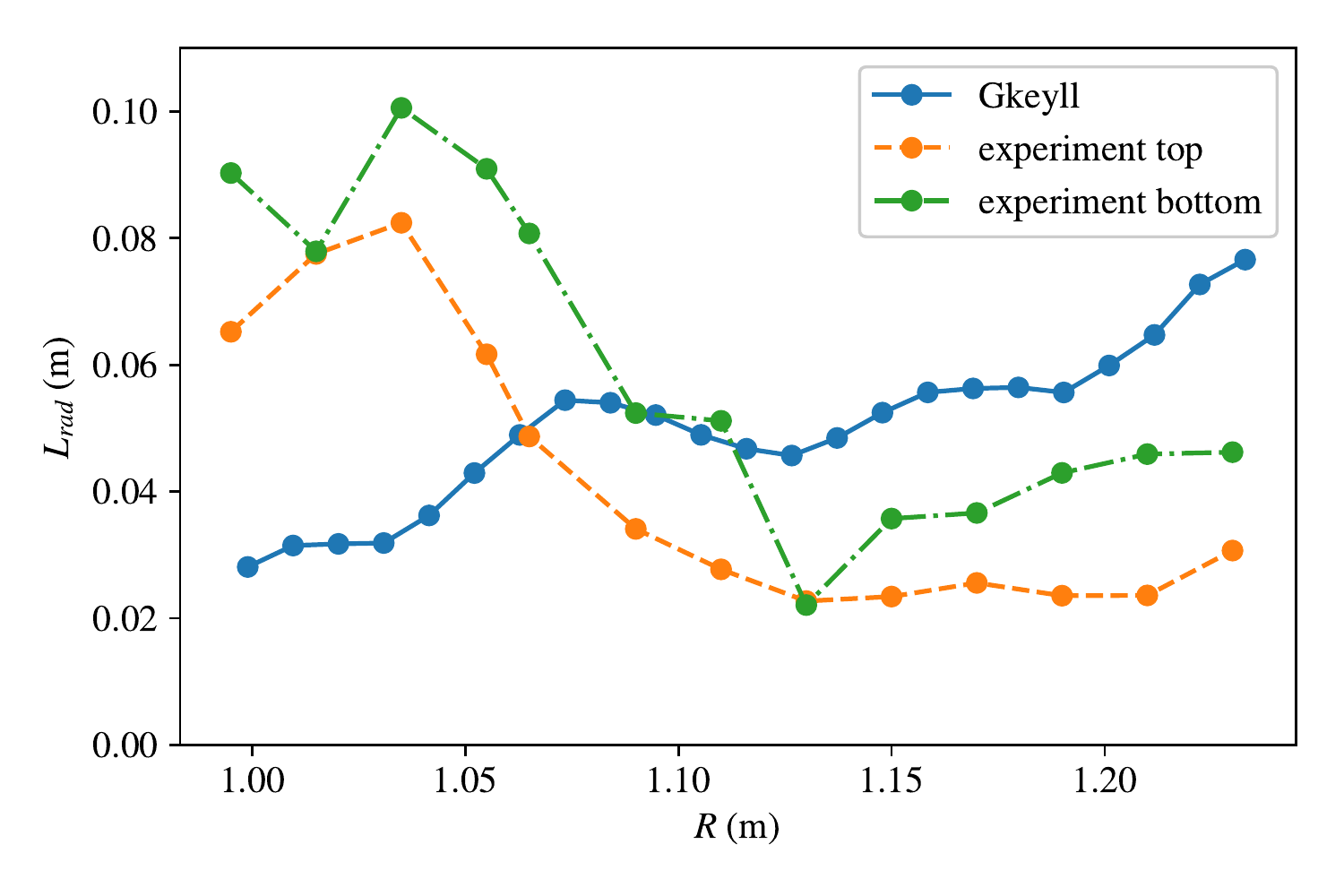}
    \caption{Comparison of radial correlation lengths $L_{\mathrm{rad}}$ as a function of radius, from the simulation and from measurements at the top and bottom of the experiment.}
    \label{fig:lrad}
\end{figure}
\begin{figure}
    \includegraphics[width=0.45\textwidth]{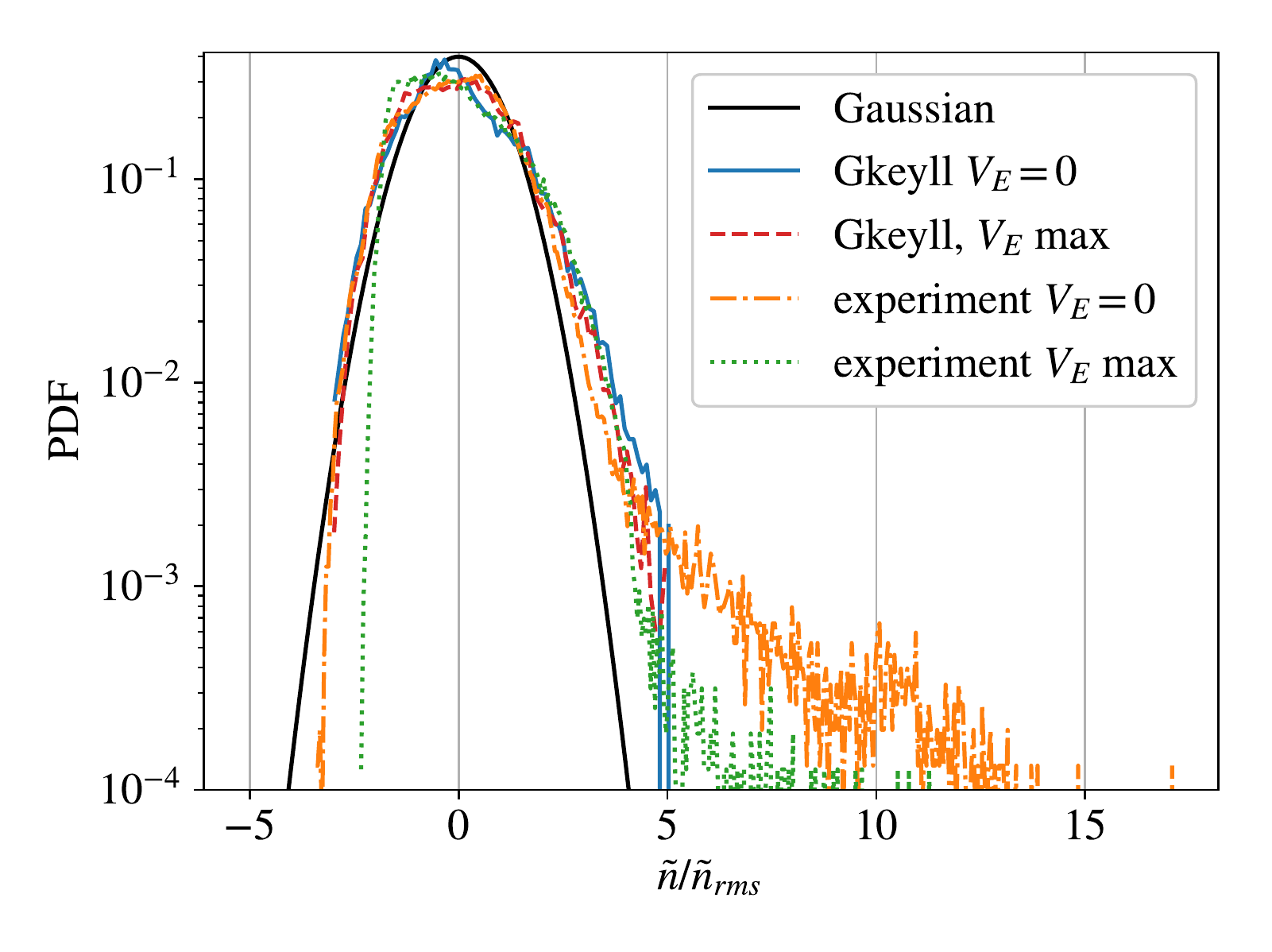}
    \caption{Probability density functions for density fluctuations normalized to the root-mean-square are compared to a Gaussian distribution. Experimental and simulation data are compared where $V_E$ is zero and maximal. All curves show non-Gaussian features, but the simulation did not capture the large positive density fluctuations of the experiment in the given run-time.}
    \label{fig:pdf}
\end{figure}
\begin{figure}
    \includegraphics[width=0.5\textwidth]{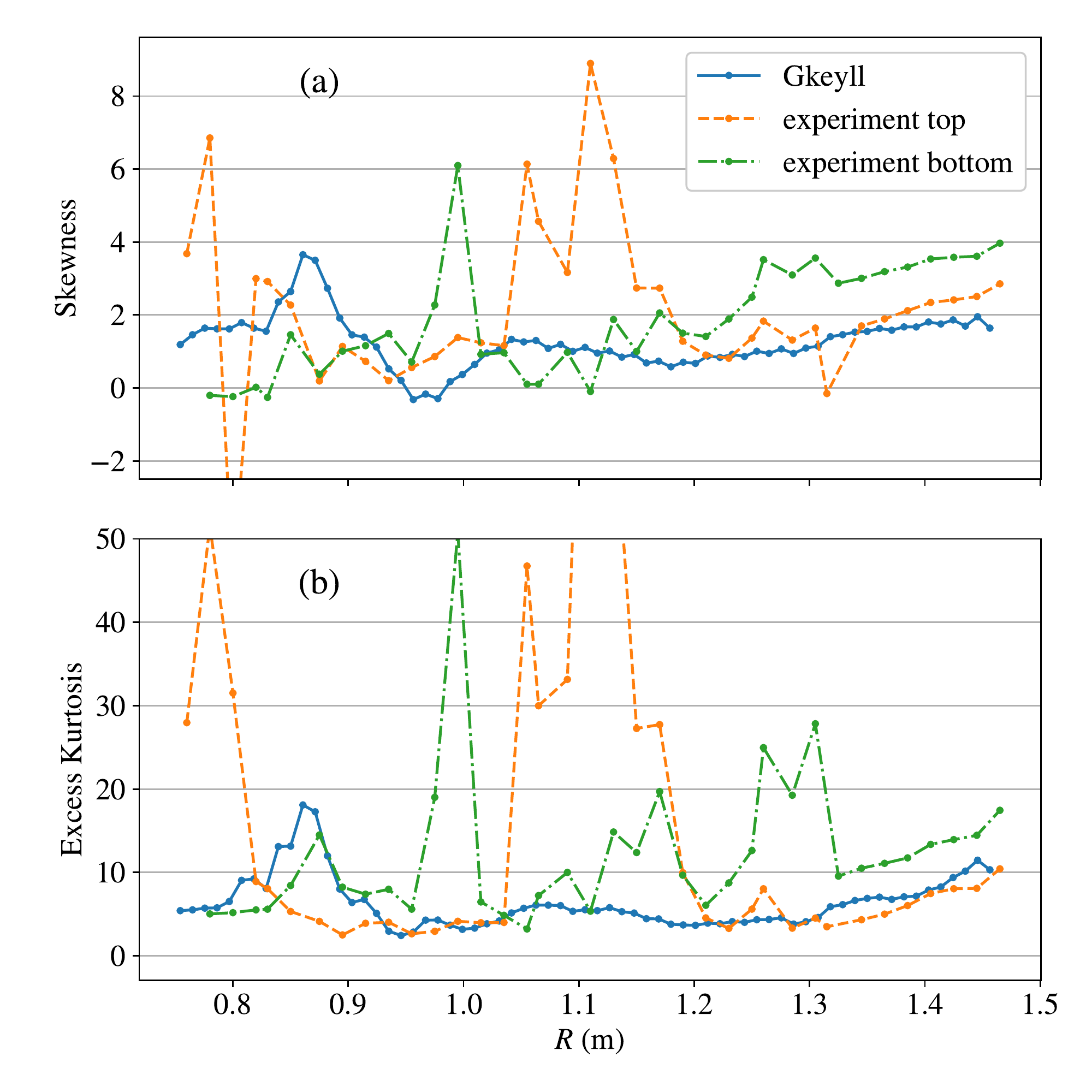}
    \caption{Comparison of (a) skewness and (b) excess kurtosis of density fluctuations as a function of radius for simulation and experimental values.}
    \label{fig:skew-kurt}
\end{figure}
\begin{figure*}
    \includegraphics[width=0.9\textwidth]{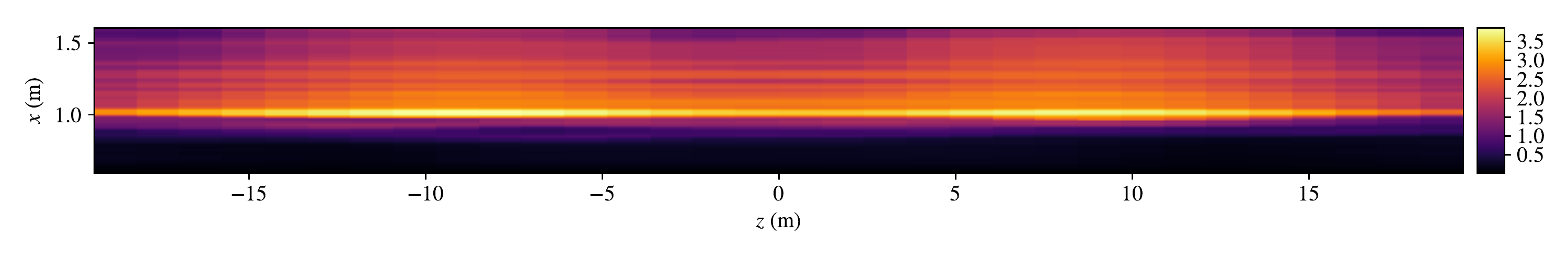}
    \caption{A snapshot of electron density $n_e(x,y=0,z)$ at the midpoint in $y$ at 10 ms. The aspect ratio is not to scale.}
    \label{fig:xz}
\end{figure*}
\begin{figure}
    \includegraphics[width=0.5\textwidth]{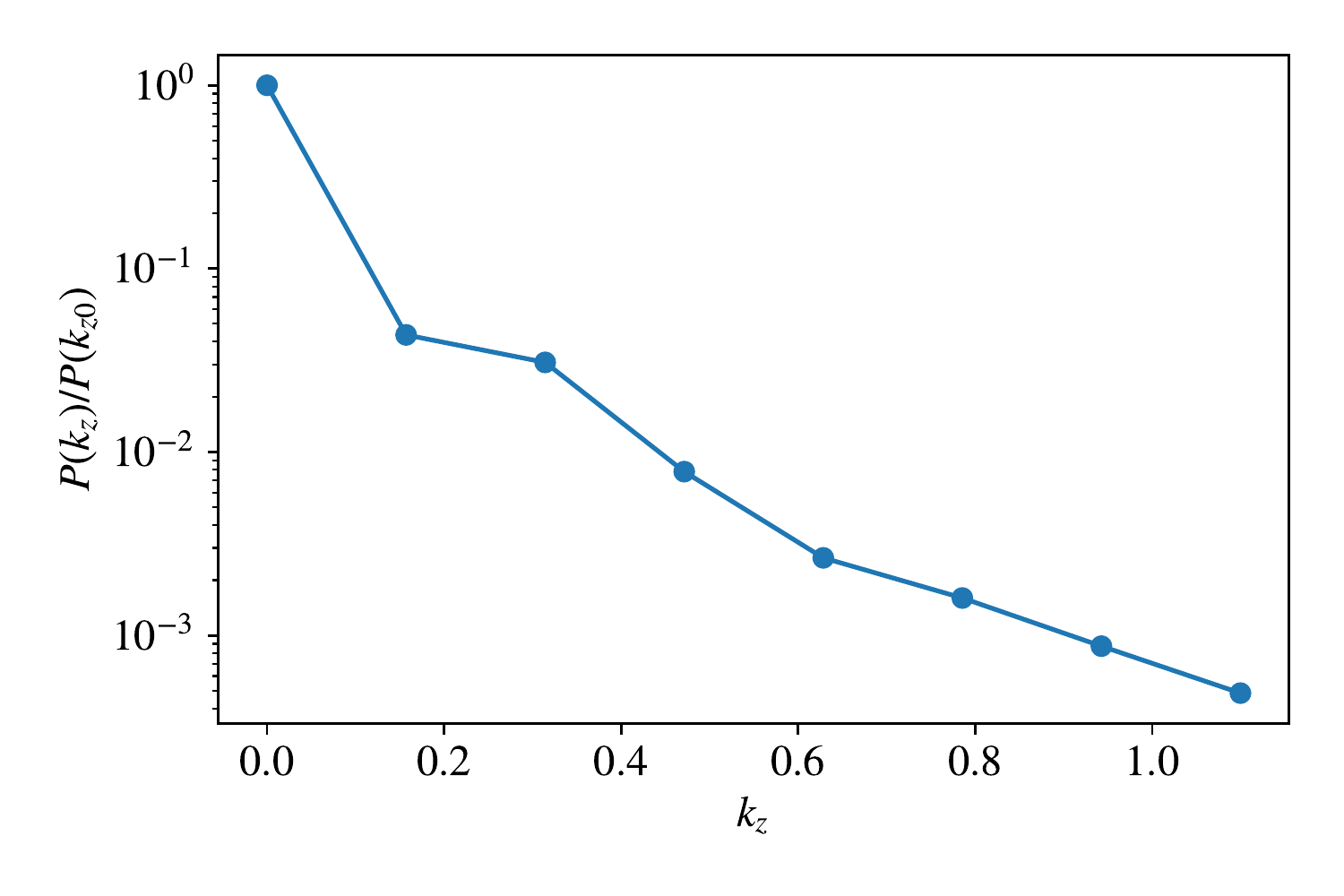}
    \caption{Power in the Fourier transform in $z$ of electron density fluctuations, averaged in $x$ and $y$ and from 10 to 16 ms, from simulation data. The power in the zeroth mode is at least ten times greater than the other modes, suggesting the average $k_z \approx 0$.}
    \label{fig:kz}
\end{figure}  
\indent Figure \ref{fig:autocorr} depicts the autocorrelation functions, which are calculated as $C(\tau) = \langle \tilde{n}(t) \tilde{n}(t+\tau) \rangle / \langle \tilde{n}(t)^2 \rangle$. Brackets denote a time average, and $\tau$ is the time difference. In both sets of data, the curves where $V_E$ is maximal exhibit more oscillatory behavior. The correlation time for the simulation data is shorter, as we would expect given the results from the power spectra. \\ 
\indent Figure \ref{fig:lrad} compares the radial correlation length as a function of radius. The experimental and simulation graphs show opposite trends, though the magnitude is similar. Reduced turbulence levels have been shown to correspond to a reduction in the radial correlation length.\cite{gentle2010comparison,terry2000suppression} However, the correlation lengths are larger for the simulation at $R>1.1$~m, though the turbulence fluctuations are less than experiment in this region. \\
\indent Figure \ref{fig:pdf} compares the probability density functions (PDF's) at radial locations where $V_E$ is zero and maximal. Experimental values were measured at the top of the device and calculated as $\tilde{n}/\tilde{n}_{\text{rms}}$. The positive tails of the simulation PDF's approach the experimental values until 5$\tilde{n}/\tilde{n}_{rms}$, where they end. Presumably, if the simulations were run for a longer time, the tail in the simulation PDF would extend to higher $\tilde{n}/\tilde{n}_{\rm rms}$. For radii where $V_E$ is maximal, negative tails in the simulation are greater than experiment. The simulations have lower density fluctuations at large radii, compared with experiment, which means that the negative fluctuations in density are less affected by the positivity of total density and could possibly explain this difference in the negative tails. \\
\indent In figure \ref{fig:skew-kurt}(a) experimental and simulation density fluctuations exhibit positive skewness, which is a characteristic of blob transport.\cite{d2011convective} Skewness is calculated as $E[\tilde{n}^3]/\sigma^3$, where $E[...]$ is the expectation value and $\sigma$ is the standard deviation of the density fluctuations. There are similarities between the simulation and experimental measurements from the top at radii greater than 1.2 m. However, the simulation does not reproduce the large values observed in the experiment at $R=1.1$ m for the top measurements and at $R=1.0$ for the bottom measurements. We compare excess kurtosis in Fig.~\ref{fig:skew-kurt}(b), which is calculated as $E[\tilde{n}^4]/\sigma^4 - 3$. Positive excess kurtosis is also a signature of blob transport.\cite{d2011convective} Though the simulation matches the top experimental curve at radii greater than 1.25 m, it fails to capture the large peaks in excess kurtosis that appear in both experimental data sets. We also note that values of skewness and kurtosis from Gkeyll simulations with NSTX-like parameters are comparable to those calculated here.\cite{Shi2017thesis,shi2019full} \\
\indent Figure \ref{fig:xz} shows a plot of the electron density $(x,z)$ at the midpoint in $y$, and there is little structure along $z$ in the bad curvature region $R > 1.0$ m, except in the overall fall-off of density approaching the conducting-sheath boundaries. This suggests interchange-like turbulence, with $k_z \approx 0$. Figure \ref{fig:kz} depicts the power in the Fourier transform of the density fluctuations, $\tilde{n} = n - \langle n \rangle_y$, in $z$, averaged in $x$ and $y$ and in time from 10 to 16 ms. The zeroth mode is more than an order of magnitude greater than the other modes, suggesting that the average $k_z \approx 0$ and, hence, interchange-driven turbulence dominates. Limitations in computational resources prevented high-resolution comparisons with lower pitch angles, particularly because the parallel ion transit times are longer and require longer simulation times to saturate. Low-resolution test cases demonstrated greater $k_z$ structure as the pitch angle was decreased, possibly indicating a transition to the drift-interchange-driven turbulence regime as predicted in Refs.~\citenum{poli2008transition} and \citenum{ricci2010turbulence}.
\begin{figure}
    \includegraphics[width=0.5\textwidth]{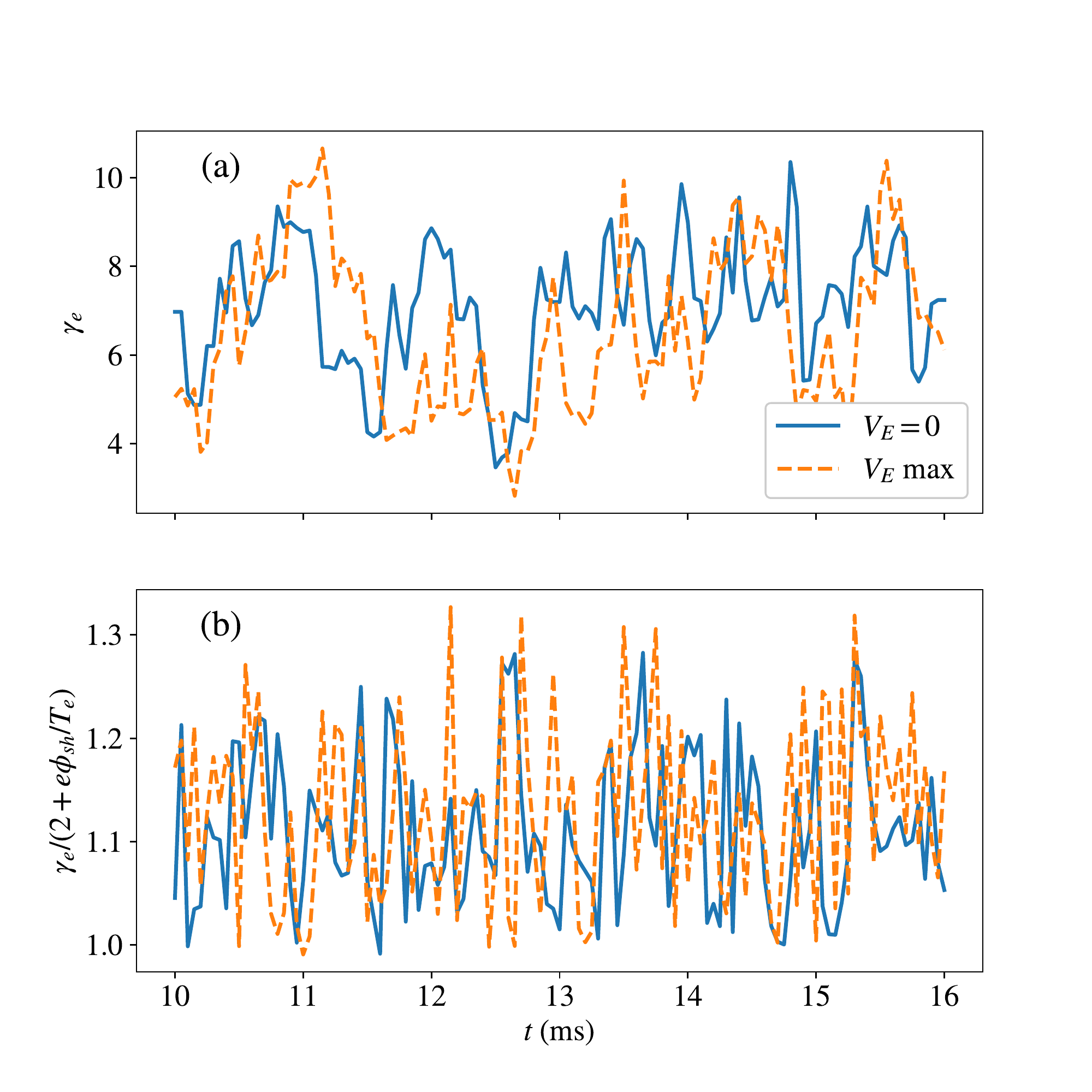}
    \caption{Time evolution of electron sheath heat-transmission coefficient at radial locations where $V_E$ is zero and maximal. In (a), the coefficient is given by $\gamma_e = q_\parallel/(\Gamma_{\parallel,n} T_e)$, with quantities evaluated at the sheath entrance, $z=L_z/2$, and at the midpoint in $y$. In (b), $\gamma_e$ is normalized to the expected value for a Maxwellian distribution function, $2 + e\phi_{\rm sh}/T_e$, which shows $\gamma_e$ fluctuating in time and exceeding the expected Maxwellian value by approximately 10--20\%, indicating relatively small kinetic effects.}
    \label{fig:gamma_e}
\end{figure}  

One way kinetic effects enter is via the fact that collisions that are not strong enough to fully thermalize electrons in the tail above the sheath potential cutoff velocity during a single bounce of electrons between the sheaths at the two ends of the field line. In some places this may reduce the electron tail and reduce the electron sheath heat-transmission coefficient, but in places where turbulence has carried energetic electrons from a hot region to a region that is colder on average, it may enhance the electron sheath heat-transmission coefficient. The coefficient is given by $\gamma_e = q_\parallel/(\Gamma_{\parallel,n} T_e)$, with quantities evaluated at the sheath entrance, $z=L_z/2$, and at the midpoint in $y$. We plot the time-evolution of this quantity in figure \ref{fig:gamma_e}(a) for radial values where $V_E$ is zero and maximal. For a Maxwellian distribution function, this coefficient is $2 + e\phi_{\rm sh}/T_e$, and it is often assumed $e \phi_{\rm sh}/T_e \sim 3$.\cite{stangeby2000plasma} In Fig.~\ref{fig:gamma_e}(b), we normalize the electron sheath heat-transmission coefficient to this quantity, using the time-dependent quantities $\phi_{\rm sh}$ and $T_e$ rather than assuming $e \phi_{\rm sh}/T_e=3$. In the simulation, the sheath heat-transmission coefficient relative to the Maxwellian value fluctuates in time, exceeding the expected Maxwellian coefficient by approximately 10--20\%, suggesting relatively small, but non-negligible kinetic effects for these parameters. However, a more detailed convergence study in velocity resolution would be needed to quantify how much of this is due to numerical discretization versus kinetic effects. 

Kinetic effects may be much more important in other regimes, such as in some previous PIC simulations of ELM-type events\cite{Pitts_2007} on JET, where they observed that $\gamma_e$ was 5 to 6 times the nominal Maxwellian value. We leave more detailed analysis of kinetic effects to future simulations of such regimes and also when we extend the code, as planned, to handle the closed-field-line pedestal and SOL regions of tokamaks simultaneously.

\section{Conclusion} \label{sec:conclusion}
\indent This work represents the first extensive comparison of kinetic simulations of a toroidal open-field line basic plasma physics device with experimental data. The results demonstrate good progress towards modeling turbulence on helical open-field lines in tokamak SOL-like conditions with gyrokinetic equations. These results indicate which aspects of the model could be improved prior to simulating a more complicated tokamak SOL where kinetic effects are likely more significant. \\
\indent Simulation density profiles approached Helimak experimental values of magnitude and gradient scale length but failed to caputure the top-bottom asymmetry. The simulation also under-predicted the magnitude and gradient of the temperature profiles. Turbulence fluctuation levels were closer to experiment than previous fluid simulations, particularly near $R=1$ m, but still less than experiment at higher radii. Some qualitative features of turbulence statistics were captured, such as oscillations in the autocorrelation function where $V_E$ is maximal, radial correlation lengths of similar magnitude, and positive skewness and excess kurtosis of the distribution function in the bad-curvature region. Notable differences were that the power spectra from the simulation peaked at higher frequencies than experiment, radial correlation values have opposing trends, and the simulation did not capture the largest values of skewness and excess kurtosis at some experimental radial locations. Previous research suggests that interchange turbulence should dominate for this connection length, and this was confirmed with an average $k_\parallel \approx 0$ in our simulations. Kinetic effects were observed in fluctuations of the electron sheath heat-transmission coefficient.\\
\indent It is clear that key physical and geometric features need to be added to our model in the future for better overall agreement with experimental turbulence statistics. A more self-consistent source model that includes radiation and the density-dependent upper-hybrid resonance might improve the agreement of equilibrium profiles. Including a vertical ($z$-directed, in our non-orthogonal coordinates) \ExB\ flow may account for some of the top-bottom asymmetry that is visible in the density equilibrium profiles. Since the experimental background density varies radially, a more accurate model would use the density-weighted Poisson equation (instead of \ref{eq:poisson}), which includes the time and spatial variation in the ion guiding-center density on the left-hand side of the equation. Other geometric features to include are magnetic shear and the radial variation in the connection length. These may affect where the transition from interchange to drift-wave turbulence occurs by giving the turbulence shorter parallel correlation length and breaking up the strong oscillatory dynamics present in the simulation. \\
\indent A new, faster version of the Gkeyll code is being developed and will allow for scans in the ion-to-electron mass ratio and collision frequencies to study their effect on turbulence statistics. Scans in connection length could also be performed to study the transition from interchange to drift-interchange turbulence. The effect of neutral particles is significant in this parameter regime, and future development for the code will likely include an appropriate neutral model. Improvements to the conducting-sheath boundary conditions will also be considered.
\begin{acknowledgements}
T.N.B. would like to thank F. Ebrahimi for reviewing the manuscript and for helpful discussions on turbulence spectra and convergence testing. We also thank N. Mandell for helpful discussions about non-orthogonal coordinate system. This work used the Extreme Science and Engineering Discovery Environment (XSEDE), which is supported by National Science Foundation grant number ACI-1548562. We thank D.R. Hatch, J.M. Tenbarge, and F.L. Waelbroeck for contributing computational resources through their allocations on the Texas Advanced Computing Center Stampede2 cluster. We also thank J. Juno for helping set up Gkeyll on Stampede. This work was supported by the U.S. Department of Energy SCGSR program under contract DE-SC0014664, by DOE Contract DE-FG02-04ER-54742, through the Institute of Fusion Studies at the University of Texas at Austin, and DOE contract DE-AC02-09CH11466, through the Princeton Plasma Physics Laboratory (PPPL). G.W.H. and A.H. were supported in part by the SciDAC Partnership for Multiscale Gyrokinetics and the SciDAC Partnership for High-Fidelity Boundary Plasma Simulations. A.H. was also supported in part by the PPPL Laboratory Directed Research and Development program.
\end{acknowledgements}

\bibliography{references}

\begin{thebibliography}{45}%
\makeatletter
\providecommand \@ifxundefined [1]{%
 \@ifx{#1\undefined}
}%
\providecommand \@ifnum [1]{%
 \ifnum #1\expandafter \@firstoftwo
 \else \expandafter \@secondoftwo
 \fi
}%
\providecommand \@ifx [1]{%
 \ifx #1\expandafter \@firstoftwo
 \else \expandafter \@secondoftwo
 \fi
}%
\providecommand \natexlab [1]{#1}%
\providecommand \enquote  [1]{``#1''}%
\providecommand \bibnamefont  [1]{#1}%
\providecommand \bibfnamefont [1]{#1}%
\providecommand \citenamefont [1]{#1}%
\providecommand \href@noop [0]{\@secondoftwo}%
\providecommand \href [0]{\begingroup \@sanitize@url \@href}%
\providecommand \@href[1]{\@@startlink{#1}\@@href}%
\providecommand \@@href[1]{\endgroup#1\@@endlink}%
\providecommand \@sanitize@url [0]{\catcode `\\12\catcode `\$12\catcode
  `\&12\catcode `\#12\catcode `\^12\catcode `\_12\catcode `\%12\relax}%
\providecommand \@@startlink[1]{}%
\providecommand \@@endlink[0]{}%
\providecommand \url  [0]{\begingroup\@sanitize@url \@url }%
\providecommand \@url [1]{\endgroup\@href {#1}{\urlprefix }}%
\providecommand \urlprefix  [0]{URL }%
\providecommand \Eprint [0]{\href }%
\providecommand \doibase [0]{http://dx.doi.org/}%
\providecommand \selectlanguage [0]{\@gobble}%
\providecommand \bibinfo  [0]{\@secondoftwo}%
\providecommand \bibfield  [0]{\@secondoftwo}%
\providecommand \translation [1]{[#1]}%
\providecommand \BibitemOpen [0]{}%
\providecommand \bibitemStop [0]{}%
\providecommand \bibitemNoStop [0]{.\EOS\space}%
\providecommand \EOS [0]{\spacefactor3000\relax}%
\providecommand \BibitemShut  [1]{\csname bibitem#1\endcsname}%
\let\auto@bib@innerbib\@empty
\bibitem [{\citenamefont {Eich}\ \emph {et~al.}(2013)\citenamefont {Eich},
  \citenamefont {Leonard}, \citenamefont {Pitts}, \citenamefont {Fundamenski},
  \citenamefont {Goldston}, \citenamefont {Gray}, \citenamefont {Herrmann},
  \citenamefont {Kirk}, \citenamefont {Kallenbach}, \citenamefont {Kardaun},
  \citenamefont {Kukushkin}, \citenamefont {LaBombard}, \citenamefont {Maingi},
  \citenamefont {Makowski}, \citenamefont {Scarabosio}, \citenamefont
  {Sieglin}, \citenamefont {Terry}, \citenamefont {Thornton}, \citenamefont
  {{ASDEX Upgrade Team}},\ and\ \citenamefont {{JET EFDA
  Contributors}}}]{Eich2013}%
  \BibitemOpen
  \bibfield  {author} {\bibinfo {author} {\bibfnamefont {T.}~\bibnamefont
  {Eich}}, \bibinfo {author} {\bibfnamefont {A.}~\bibnamefont {Leonard}},
  \bibinfo {author} {\bibfnamefont {R.}~\bibnamefont {Pitts}}, \bibinfo
  {author} {\bibfnamefont {W.}~\bibnamefont {Fundamenski}}, \bibinfo {author}
  {\bibfnamefont {R.}~\bibnamefont {Goldston}}, \bibinfo {author}
  {\bibfnamefont {T.}~\bibnamefont {Gray}}, \bibinfo {author} {\bibfnamefont
  {A.}~\bibnamefont {Herrmann}}, \bibinfo {author} {\bibfnamefont
  {A.}~\bibnamefont {Kirk}}, \bibinfo {author} {\bibfnamefont {A.}~\bibnamefont
  {Kallenbach}}, \bibinfo {author} {\bibfnamefont {O.}~\bibnamefont {Kardaun}},
  \bibinfo {author} {\bibfnamefont {A.}~\bibnamefont {Kukushkin}}, \bibinfo
  {author} {\bibfnamefont {B.}~\bibnamefont {LaBombard}}, \bibinfo {author}
  {\bibfnamefont {R.}~\bibnamefont {Maingi}}, \bibinfo {author} {\bibfnamefont
  {M.}~\bibnamefont {Makowski}}, \bibinfo {author} {\bibfnamefont
  {A.}~\bibnamefont {Scarabosio}}, \bibinfo {author} {\bibfnamefont
  {B.}~\bibnamefont {Sieglin}}, \bibinfo {author} {\bibfnamefont
  {J.}~\bibnamefont {Terry}}, \bibinfo {author} {\bibfnamefont
  {A.}~\bibnamefont {Thornton}}, \bibinfo {author} {\bibnamefont {{ASDEX
  Upgrade Team}}}, \ and\ \bibinfo {author} {\bibnamefont {{JET EFDA
  Contributors}}},\ }\href {http://stacks.iop.org/0029-5515/53/i=9/a=093031}
  {\bibfield  {journal} {\bibinfo  {journal} {Nucl. Fusion}\ }\textbf {\bibinfo
  {volume} {53}},\ \bibinfo {pages} {093031} (\bibinfo {year}
  {2013})}\BibitemShut {NoStop}%
\bibitem [{\citenamefont {Goldston}(2012)}]{Goldston:2012}%
  \BibitemOpen
  \bibfield  {author} {\bibinfo {author} {\bibfnamefont {R.}~\bibnamefont
  {Goldston}},\ }\href {http://stacks.iop.org/0029-5515/52/i=1/a=013009}
  {\bibfield  {journal} {\bibinfo  {journal} {Nucl. Fusion}\ }\textbf {\bibinfo
  {volume} {52}},\ \bibinfo {pages} {013009} (\bibinfo {year}
  {2012})}\BibitemShut {NoStop}%
\bibitem [{\citenamefont {Krasheninnikov}(2001)}]{krasheninnikov2001scrape}%
  \BibitemOpen
  \bibfield  {author} {\bibinfo {author} {\bibfnamefont {S.~I.}\ \bibnamefont
  {Krasheninnikov}},\ }\href
  {https://www.sciencedirect.com/science/article/pii/S0375960101002523}
  {\bibfield  {journal} {\bibinfo  {journal} {Phys. Lett. A}\ }\textbf
  {\bibinfo {volume} {283}},\ \bibinfo {pages} {368} (\bibinfo {year}
  {2001})}\BibitemShut {NoStop}%
\bibitem [{\citenamefont {Krasheninnikov}\ \emph {et~al.}(2008)\citenamefont
  {Krasheninnikov}, \citenamefont {D'Ippolito},\ and\ \citenamefont
  {Myra}}]{krasheninnikov2008recent}%
  \BibitemOpen
  \bibfield  {author} {\bibinfo {author} {\bibfnamefont {S.}~\bibnamefont
  {Krasheninnikov}}, \bibinfo {author} {\bibfnamefont {D.}~\bibnamefont
  {D'Ippolito}}, \ and\ \bibinfo {author} {\bibfnamefont {J.}~\bibnamefont
  {Myra}},\ }\href {https://doi.org/10.1017/S0022377807006940} {\bibfield
  {journal} {\bibinfo  {journal} {J. Plasma Phys.}\ }\textbf {\bibinfo {volume}
  {74}},\ \bibinfo {pages} {679} (\bibinfo {year} {2008})}\BibitemShut
  {NoStop}%
\bibitem [{\citenamefont {Zweben}\ and\ \citenamefont
  {Gould}(1985)}]{zweben1985structure}%
  \BibitemOpen
  \bibfield  {author} {\bibinfo {author} {\bibfnamefont {S.}~\bibnamefont
  {Zweben}}\ and\ \bibinfo {author} {\bibfnamefont {R.}~\bibnamefont {Gould}},\
  }\href {http://iopscience.iop.org/article/10.1088/0029-5515/25/2/005/meta}
  {\bibfield  {journal} {\bibinfo  {journal} {Nucl. Fusion}\ }\textbf {\bibinfo
  {volume} {25}},\ \bibinfo {pages} {171} (\bibinfo {year} {1985})}\BibitemShut
  {NoStop}%
\bibitem [{\citenamefont {Zweben}(1985)}]{zweben1985search}%
  \BibitemOpen
  \bibfield  {author} {\bibinfo {author} {\bibfnamefont {S.}~\bibnamefont
  {Zweben}},\ }\href {https://aip.scitation.org/doi/abs/10.1063/1.865069}
  {\bibfield  {journal} {\bibinfo  {journal} {Phys. Fluids}\ }\textbf {\bibinfo
  {volume} {28}},\ \bibinfo {pages} {974} (\bibinfo {year} {1985})}\BibitemShut
  {NoStop}%
\bibitem [{\citenamefont {Zweben}\ \emph {et~al.}(2003)\citenamefont {Zweben},
  \citenamefont {Maqueda}, \citenamefont {Stotler}, \citenamefont {Keesee},
  \citenamefont {Boedo}, \citenamefont {Bush}, \citenamefont {Kaye},
  \citenamefont {LeBlanc}, \citenamefont {Lowrance}, \citenamefont {Mastrocola}
  \emph {et~al.}}]{zweben2003high}%
  \BibitemOpen
  \bibfield  {author} {\bibinfo {author} {\bibfnamefont {S.}~\bibnamefont
  {Zweben}}, \bibinfo {author} {\bibfnamefont {R.}~\bibnamefont {Maqueda}},
  \bibinfo {author} {\bibfnamefont {D.}~\bibnamefont {Stotler}}, \bibinfo
  {author} {\bibfnamefont {A.}~\bibnamefont {Keesee}}, \bibinfo {author}
  {\bibfnamefont {J.}~\bibnamefont {Boedo}}, \bibinfo {author} {\bibfnamefont
  {C.}~\bibnamefont {Bush}}, \bibinfo {author} {\bibfnamefont {S.}~\bibnamefont
  {Kaye}}, \bibinfo {author} {\bibfnamefont {B.}~\bibnamefont {LeBlanc}},
  \bibinfo {author} {\bibfnamefont {J.}~\bibnamefont {Lowrance}}, \bibinfo
  {author} {\bibfnamefont {V.}~\bibnamefont {Mastrocola}},  \emph {et~al.},\
  }\href {http://iopscience.iop.org/article/10.1088/0029-5515/44/1/016/meta}
  {\bibfield  {journal} {\bibinfo  {journal} {Nucl. Fusion}\ }\textbf {\bibinfo
  {volume} {44}},\ \bibinfo {pages} {134} (\bibinfo {year} {2003})}\BibitemShut
  {NoStop}%
\bibitem [{\citenamefont {Terry}\ \emph {et~al.}(2007)\citenamefont {Terry},
  \citenamefont {LaBombard}, \citenamefont {Lipschultz}, \citenamefont
  {Greenwald}, \citenamefont {Rice},\ and\ \citenamefont
  {Zweben}}]{terry2007scrape}%
  \BibitemOpen
  \bibfield  {author} {\bibinfo {author} {\bibfnamefont {J.}~\bibnamefont
  {Terry}}, \bibinfo {author} {\bibfnamefont {B.}~\bibnamefont {LaBombard}},
  \bibinfo {author} {\bibfnamefont {B.}~\bibnamefont {Lipschultz}}, \bibinfo
  {author} {\bibfnamefont {M.}~\bibnamefont {Greenwald}}, \bibinfo {author}
  {\bibfnamefont {J.}~\bibnamefont {Rice}}, \ and\ \bibinfo {author}
  {\bibfnamefont {S.}~\bibnamefont {Zweben}},\ }\href
  {https://www.tandfonline.com/doi/abs/10.13182/FST07-A1426} {\bibfield
  {journal} {\bibinfo  {journal} {Fusion. Sci. Tech.}\ }\textbf {\bibinfo
  {volume} {51}},\ \bibinfo {pages} {342} (\bibinfo {year} {2007})}\BibitemShut
  {NoStop}%
\bibitem [{\citenamefont {Boedo}\ \emph {et~al.}(2014)\citenamefont {Boedo},
  \citenamefont {Myra}, \citenamefont {Zweben}, \citenamefont {Maingi},
  \citenamefont {Maqueda}, \citenamefont {Soukhanovskii}, \citenamefont {Ahn},
  \citenamefont {Canik}, \citenamefont {Crocker}, \citenamefont {D'Ippolito}
  \emph {et~al.}}]{boedo2014edge}%
  \BibitemOpen
  \bibfield  {author} {\bibinfo {author} {\bibfnamefont {J.}~\bibnamefont
  {Boedo}}, \bibinfo {author} {\bibfnamefont {J.}~\bibnamefont {Myra}},
  \bibinfo {author} {\bibfnamefont {S.}~\bibnamefont {Zweben}}, \bibinfo
  {author} {\bibfnamefont {R.}~\bibnamefont {Maingi}}, \bibinfo {author}
  {\bibfnamefont {R.}~\bibnamefont {Maqueda}}, \bibinfo {author} {\bibfnamefont
  {V.}~\bibnamefont {Soukhanovskii}}, \bibinfo {author} {\bibfnamefont
  {J.}~\bibnamefont {Ahn}}, \bibinfo {author} {\bibfnamefont {J.}~\bibnamefont
  {Canik}}, \bibinfo {author} {\bibfnamefont {N.}~\bibnamefont {Crocker}},
  \bibinfo {author} {\bibfnamefont {D.}~\bibnamefont {D'Ippolito}},  \emph
  {et~al.},\ }\href {https://aip.scitation.org/doi/abs/10.1063/1.4873390}
  {\bibfield  {journal} {\bibinfo  {journal} {Phys. Plasmas}\ }\textbf
  {\bibinfo {volume} {21}},\ \bibinfo {pages} {042309} (\bibinfo {year}
  {2014})}\BibitemShut {NoStop}%
\bibitem [{\citenamefont {Stangeby}(2000)}]{stangeby2000plasma}%
  \BibitemOpen
  \bibfield  {author} {\bibinfo {author} {\bibfnamefont {P.~C.}\ \bibnamefont
  {Stangeby}},\ }\href@noop {} {\emph {\bibinfo {title} {The plasma boundary of
  magnetic fusion devices}}}\ (\bibinfo  {publisher} {CRC Press},\ \bibinfo
  {year} {2000})\BibitemShut {NoStop}%
\bibitem [{\citenamefont {Shimada}\ \emph {et~al.}(2007)\citenamefont
  {Shimada}, \citenamefont {Campbell}, \citenamefont {Mukhovatov},
  \citenamefont {Fujiwara}, \citenamefont {Kirneva}, \citenamefont {Lackner},
  \citenamefont {Nagami}, \citenamefont {Pustovitov}, \citenamefont {Uckan},
  \citenamefont {Wesley}, \citenamefont {Asakura}, \citenamefont {Costley},
  \citenamefont {Donné}, \citenamefont {Doyle}, \citenamefont {Fasoli},
  \citenamefont {Gormezano}, \citenamefont {Gribov}, \citenamefont {Gruber},
  \citenamefont {Hender}, \citenamefont {Houlberg}, \citenamefont {Ide},
  \citenamefont {Kamada}, \citenamefont {Leonard}, \citenamefont {Lipschultz},
  \citenamefont {Loarte}, \citenamefont {Miyamoto}, \citenamefont {Mukhovatov},
  \citenamefont {Osborne}, \citenamefont {Polevoi},\ and\ \citenamefont
  {Sips}}]{Shimada2007}%
  \BibitemOpen
  \bibfield  {author} {\bibinfo {author} {\bibfnamefont {M.}~\bibnamefont
  {Shimada}}, \bibinfo {author} {\bibfnamefont {D.}~\bibnamefont {Campbell}},
  \bibinfo {author} {\bibfnamefont {V.}~\bibnamefont {Mukhovatov}}, \bibinfo
  {author} {\bibfnamefont {M.}~\bibnamefont {Fujiwara}}, \bibinfo {author}
  {\bibfnamefont {N.}~\bibnamefont {Kirneva}}, \bibinfo {author} {\bibfnamefont
  {K.}~\bibnamefont {Lackner}}, \bibinfo {author} {\bibfnamefont
  {M.}~\bibnamefont {Nagami}}, \bibinfo {author} {\bibfnamefont
  {V.}~\bibnamefont {Pustovitov}}, \bibinfo {author} {\bibfnamefont
  {N.}~\bibnamefont {Uckan}}, \bibinfo {author} {\bibfnamefont
  {J.}~\bibnamefont {Wesley}}, \bibinfo {author} {\bibfnamefont
  {N.}~\bibnamefont {Asakura}}, \bibinfo {author} {\bibfnamefont
  {A.}~\bibnamefont {Costley}}, \bibinfo {author} {\bibfnamefont
  {A.}~\bibnamefont {Donné}}, \bibinfo {author} {\bibfnamefont
  {E.}~\bibnamefont {Doyle}}, \bibinfo {author} {\bibfnamefont
  {A.}~\bibnamefont {Fasoli}}, \bibinfo {author} {\bibfnamefont
  {C.}~\bibnamefont {Gormezano}}, \bibinfo {author} {\bibfnamefont
  {Y.}~\bibnamefont {Gribov}}, \bibinfo {author} {\bibfnamefont
  {O.}~\bibnamefont {Gruber}}, \bibinfo {author} {\bibfnamefont
  {T.}~\bibnamefont {Hender}}, \bibinfo {author} {\bibfnamefont
  {W.}~\bibnamefont {Houlberg}}, \bibinfo {author} {\bibfnamefont
  {S.}~\bibnamefont {Ide}}, \bibinfo {author} {\bibfnamefont {Y.}~\bibnamefont
  {Kamada}}, \bibinfo {author} {\bibfnamefont {A.}~\bibnamefont {Leonard}},
  \bibinfo {author} {\bibfnamefont {B.}~\bibnamefont {Lipschultz}}, \bibinfo
  {author} {\bibfnamefont {A.}~\bibnamefont {Loarte}}, \bibinfo {author}
  {\bibfnamefont {K.}~\bibnamefont {Miyamoto}}, \bibinfo {author}
  {\bibfnamefont {V.}~\bibnamefont {Mukhovatov}}, \bibinfo {author}
  {\bibfnamefont {T.}~\bibnamefont {Osborne}}, \bibinfo {author} {\bibfnamefont
  {A.}~\bibnamefont {Polevoi}}, \ and\ \bibinfo {author} {\bibfnamefont
  {A.}~\bibnamefont {Sips}},\ }\href {\doibase 10.1088/0029-5515/47/6/S01}
  {\bibfield  {journal} {\bibinfo  {journal} {Nucl. Fusion}\ }\textbf {\bibinfo
  {volume} {47}},\ \bibinfo {pages} {S1} (\bibinfo {year} {2007})}\BibitemShut
  {NoStop}%
\bibitem [{\citenamefont {Loarte}\ \emph {et~al.}(2007)\citenamefont {Loarte},
  \citenamefont {Lipschultz}, \citenamefont {Kukushkin}, \citenamefont
  {Matthews}, \citenamefont {Stangeby}, \citenamefont {Asakura}, \citenamefont
  {Counsell}, \citenamefont {Federici}, \citenamefont {Kallenbach},
  \citenamefont {Krieger} \emph {et~al.}}]{loarte2007power}%
  \BibitemOpen
  \bibfield  {author} {\bibinfo {author} {\bibfnamefont {A.}~\bibnamefont
  {Loarte}}, \bibinfo {author} {\bibfnamefont {B.}~\bibnamefont {Lipschultz}},
  \bibinfo {author} {\bibfnamefont {A.}~\bibnamefont {Kukushkin}}, \bibinfo
  {author} {\bibfnamefont {G.}~\bibnamefont {Matthews}}, \bibinfo {author}
  {\bibfnamefont {P.}~\bibnamefont {Stangeby}}, \bibinfo {author}
  {\bibfnamefont {N.}~\bibnamefont {Asakura}}, \bibinfo {author} {\bibfnamefont
  {G.}~\bibnamefont {Counsell}}, \bibinfo {author} {\bibfnamefont
  {G.}~\bibnamefont {Federici}}, \bibinfo {author} {\bibfnamefont
  {A.}~\bibnamefont {Kallenbach}}, \bibinfo {author} {\bibfnamefont
  {K.}~\bibnamefont {Krieger}},  \emph {et~al.},\ }\href
  {http://iopscience.iop.org/article/10.1088/0029-5515/47/6/S04/meta}
  {\bibfield  {journal} {\bibinfo  {journal} {Nucl. Fusion}\ }\textbf {\bibinfo
  {volume} {47}},\ \bibinfo {pages} {S203} (\bibinfo {year}
  {2007})}\BibitemShut {NoStop}%
\bibitem [{\citenamefont {Gentle}\ and\ \citenamefont
  {He}(2008)}]{gentle2008texas}%
  \BibitemOpen
  \bibfield  {author} {\bibinfo {author} {\bibfnamefont {K.~W.}\ \bibnamefont
  {Gentle}}\ and\ \bibinfo {author} {\bibfnamefont {H.}~\bibnamefont {He}},\
  }\href {http://iopscience.iop.org/article/10.1088/1009-0630/10/3/03/meta}
  {\bibfield  {journal} {\bibinfo  {journal} {Plasma Sci. Tech.}\ }\textbf
  {\bibinfo {volume} {10}},\ \bibinfo {pages} {284} (\bibinfo {year}
  {2008})}\BibitemShut {NoStop}%
\bibitem [{\citenamefont {Perez}\ \emph {et~al.}(2006)\citenamefont {Perez},
  \citenamefont {Horton}, \citenamefont {Gentle}, \citenamefont {Rowan},
  \citenamefont {Lee},\ and\ \citenamefont {Dahlburg}}]{perez2006drift}%
  \BibitemOpen
  \bibfield  {author} {\bibinfo {author} {\bibfnamefont {J.~C.}\ \bibnamefont
  {Perez}}, \bibinfo {author} {\bibfnamefont {W.}~\bibnamefont {Horton}},
  \bibinfo {author} {\bibfnamefont {K.}~\bibnamefont {Gentle}}, \bibinfo
  {author} {\bibfnamefont {W.}~\bibnamefont {Rowan}}, \bibinfo {author}
  {\bibfnamefont {K.}~\bibnamefont {Lee}}, \ and\ \bibinfo {author}
  {\bibfnamefont {R.~B.}\ \bibnamefont {Dahlburg}},\ }\href
  {https://doi.org/10.1063/1.2168401} {\bibfield  {journal} {\bibinfo
  {journal} {Phys. Plasmas}\ }\textbf {\bibinfo {volume} {13}},\ \bibinfo
  {pages} {032101} (\bibinfo {year} {2006})}\BibitemShut {NoStop}%
\bibitem [{\citenamefont {Poli}\ \emph {et~al.}(2006)\citenamefont {Poli},
  \citenamefont {Brunner}, \citenamefont {Diallo}, \citenamefont {Fasoli},
  \citenamefont {Furno}, \citenamefont {Labit}, \citenamefont {M{\"u}ller},
  \citenamefont {Plyushchev},\ and\ \citenamefont
  {Podest{\`a}}}]{poli2006experimental}%
  \BibitemOpen
  \bibfield  {author} {\bibinfo {author} {\bibfnamefont {F.}~\bibnamefont
  {Poli}}, \bibinfo {author} {\bibfnamefont {S.}~\bibnamefont {Brunner}},
  \bibinfo {author} {\bibfnamefont {A.}~\bibnamefont {Diallo}}, \bibinfo
  {author} {\bibfnamefont {A.}~\bibnamefont {Fasoli}}, \bibinfo {author}
  {\bibfnamefont {I.}~\bibnamefont {Furno}}, \bibinfo {author} {\bibfnamefont
  {B.}~\bibnamefont {Labit}}, \bibinfo {author} {\bibfnamefont
  {S.}~\bibnamefont {M{\"u}ller}}, \bibinfo {author} {\bibfnamefont
  {G.}~\bibnamefont {Plyushchev}}, \ and\ \bibinfo {author} {\bibfnamefont
  {M.}~\bibnamefont {Podest{\`a}}},\ }\href
  {https://aip.scitation.org/doi/abs/10.1063/1.2356483} {\bibfield  {journal}
  {\bibinfo  {journal} {Phys. Plasmas}\ }\textbf {\bibinfo {volume} {13}},\
  \bibinfo {pages} {102104} (\bibinfo {year} {2006})}\BibitemShut {NoStop}%
\bibitem [{\citenamefont {Poli}\ \emph {et~al.}(2008)\citenamefont {Poli},
  \citenamefont {Ricci}, \citenamefont {Fasoli},\ and\ \citenamefont
  {Podest{\`a}}}]{poli2008transition}%
  \BibitemOpen
  \bibfield  {author} {\bibinfo {author} {\bibfnamefont {F.}~\bibnamefont
  {Poli}}, \bibinfo {author} {\bibfnamefont {P.}~\bibnamefont {Ricci}},
  \bibinfo {author} {\bibfnamefont {A.}~\bibnamefont {Fasoli}}, \ and\ \bibinfo
  {author} {\bibfnamefont {M.}~\bibnamefont {Podest{\`a}}},\ }\href
  {https://aip.scitation.org/doi/abs/10.1063/1.2899303} {\bibfield  {journal}
  {\bibinfo  {journal} {Phys. Plasmas}\ }\textbf {\bibinfo {volume} {15}},\
  \bibinfo {pages} {032104} (\bibinfo {year} {2008})}\BibitemShut {NoStop}%
\bibitem [{\citenamefont {Ricci}\ \emph {et~al.}(2008)\citenamefont {Ricci},
  \citenamefont {Rogers},\ and\ \citenamefont {Brunner}}]{ricci2008high}%
  \BibitemOpen
  \bibfield  {author} {\bibinfo {author} {\bibfnamefont {P.}~\bibnamefont
  {Ricci}}, \bibinfo {author} {\bibfnamefont {B.}~\bibnamefont {Rogers}}, \
  and\ \bibinfo {author} {\bibfnamefont {S.}~\bibnamefont {Brunner}},\ }\href
  {https://doi.org/10.1103/PhysRevLett.100.225002} {\bibfield  {journal}
  {\bibinfo  {journal} {Phys. Rev. Lett.}\ }\textbf {\bibinfo {volume} {100}},\
  \bibinfo {pages} {225002} (\bibinfo {year} {2008})}\BibitemShut {NoStop}%
\bibitem [{\citenamefont {Williams}(2017)}]{Williams2017thesis}%
  \BibitemOpen
  \bibfield  {author} {\bibinfo {author} {\bibfnamefont {C.}~\bibnamefont
  {Williams}},\ }\emph {\bibinfo {title} {Characterization of instability
  regimes in the Helimak, a simple magnetic torus}},\ \href
  {http://hdl.handle.net/2152/47330} {Ph.D. thesis},\ \bibinfo  {school} {UT
  Austin} (\bibinfo {year} {2017})\BibitemShut {NoStop}%
\bibitem [{\citenamefont {Li}\ \emph {et~al.}(2009)\citenamefont {Li},
  \citenamefont {Rogers}, \citenamefont {Ricci},\ and\ \citenamefont
  {Gentle}}]{li2009plasma}%
  \BibitemOpen
  \bibfield  {author} {\bibinfo {author} {\bibfnamefont {B.}~\bibnamefont
  {Li}}, \bibinfo {author} {\bibfnamefont {B.}~\bibnamefont {Rogers}}, \bibinfo
  {author} {\bibfnamefont {P.}~\bibnamefont {Ricci}}, \ and\ \bibinfo {author}
  {\bibfnamefont {K.}~\bibnamefont {Gentle}},\ }\href
  {https://aip.scitation.org/doi/abs/10.1063/1.3212591} {\bibfield  {journal}
  {\bibinfo  {journal} {Phys. Plasmas}\ }\textbf {\bibinfo {volume} {16}},\
  \bibinfo {pages} {082510} (\bibinfo {year} {2009})}\BibitemShut {NoStop}%
\bibitem [{\citenamefont {Li}\ \emph {et~al.}(2011)\citenamefont {Li},
  \citenamefont {Rogers}, \citenamefont {Ricci}, \citenamefont {Gentle},\ and\
  \citenamefont {Bhattacharjee}}]{li2011turbulence}%
  \BibitemOpen
  \bibfield  {author} {\bibinfo {author} {\bibfnamefont {B.}~\bibnamefont
  {Li}}, \bibinfo {author} {\bibfnamefont {B.}~\bibnamefont {Rogers}}, \bibinfo
  {author} {\bibfnamefont {P.}~\bibnamefont {Ricci}}, \bibinfo {author}
  {\bibfnamefont {K.}~\bibnamefont {Gentle}}, \ and\ \bibinfo {author}
  {\bibfnamefont {A.}~\bibnamefont {Bhattacharjee}},\ }\href
  {https://doi.org/10.1103/PhysRevE.83.056406} {\bibfield  {journal} {\bibinfo
  {journal} {Phys. Rev. E}\ }\textbf {\bibinfo {volume} {83}},\ \bibinfo
  {pages} {056406} (\bibinfo {year} {2011})}\BibitemShut {NoStop}%
\bibitem [{\citenamefont {Ricci}\ and\ \citenamefont
  {Rogers}(2009)}]{ricci2009transport}%
  \BibitemOpen
  \bibfield  {author} {\bibinfo {author} {\bibfnamefont {P.}~\bibnamefont
  {Ricci}}\ and\ \bibinfo {author} {\bibfnamefont {B.}~\bibnamefont {Rogers}},\
  }\href {https://aip.scitation.org/doi/abs/10.1063/1.3139261} {\bibfield
  {journal} {\bibinfo  {journal} {Phys. Plasmas}\ }\textbf {\bibinfo {volume}
  {16}},\ \bibinfo {pages} {062303} (\bibinfo {year} {2009})}\BibitemShut
  {NoStop}%
\bibitem [{\citenamefont {Ricci}\ and\ \citenamefont
  {Rogers}(2010)}]{ricci2010turbulence}%
  \BibitemOpen
  \bibfield  {author} {\bibinfo {author} {\bibfnamefont {P.}~\bibnamefont
  {Ricci}}\ and\ \bibinfo {author} {\bibfnamefont {B.}~\bibnamefont {Rogers}},\
  }\href {https://doi.org/10.1103/PhysRevLett.104.145001} {\bibfield  {journal}
  {\bibinfo  {journal} {Phys. Rev. Lett.}\ }\textbf {\bibinfo {volume} {104}},\
  \bibinfo {pages} {145001} (\bibinfo {year} {2010})}\BibitemShut {NoStop}%
\bibitem [{\citenamefont {Gentle}\ \emph {et~al.}(2010)\citenamefont {Gentle},
  \citenamefont {Liao}, \citenamefont {Lee},\ and\ \citenamefont
  {Rowan}}]{gentle2010comparison}%
  \BibitemOpen
  \bibfield  {author} {\bibinfo {author} {\bibfnamefont {K.}~\bibnamefont
  {Gentle}}, \bibinfo {author} {\bibfnamefont {K.}~\bibnamefont {Liao}},
  \bibinfo {author} {\bibfnamefont {K.}~\bibnamefont {Lee}}, \ and\ \bibinfo
  {author} {\bibfnamefont {W.}~\bibnamefont {Rowan}},\ }\href
  {http://iopscience.iop.org/article/10.1088/1009-0630/12/4/02/met} {\bibfield
  {journal} {\bibinfo  {journal} {Plasma Sci. Tech.}\ }\textbf {\bibinfo
  {volume} {12}},\ \bibinfo {pages} {391} (\bibinfo {year} {2010})}\BibitemShut
  {NoStop}%
\bibitem [{\citenamefont {Gentle}\ \emph {et~al.}(2014)\citenamefont {Gentle},
  \citenamefont {Rowan}, \citenamefont {Williams},\ and\ \citenamefont
  {Brookman}}]{gentle2014turbulence}%
  \BibitemOpen
  \bibfield  {author} {\bibinfo {author} {\bibfnamefont {K.}~\bibnamefont
  {Gentle}}, \bibinfo {author} {\bibfnamefont {W.}~\bibnamefont {Rowan}},
  \bibinfo {author} {\bibfnamefont {C.}~\bibnamefont {Williams}}, \ and\
  \bibinfo {author} {\bibfnamefont {M.}~\bibnamefont {Brookman}},\ }\href
  {\doibase https://doi.org/10.1063/1.4894687} {\bibfield  {journal} {\bibinfo
  {journal} {Physics of Plasmas}\ }\textbf {\bibinfo {volume} {21}},\ \bibinfo
  {pages} {092302} (\bibinfo {year} {2014})}\BibitemShut {NoStop}%
\bibitem [{\citenamefont {Cohen}\ and\ \citenamefont
  {Xu}(2008)}]{cohen2008progress}%
  \BibitemOpen
  \bibfield  {author} {\bibinfo {author} {\bibfnamefont {R.}~\bibnamefont
  {Cohen}}\ and\ \bibinfo {author} {\bibfnamefont {X.}~\bibnamefont {Xu}},\
  }\href {https://onlinelibrary.wiley.com/doi/abs/10.1002/ctpp.200810038}
  {\bibfield  {journal} {\bibinfo  {journal} {Contrib. Plasma Phys.}\ }\textbf
  {\bibinfo {volume} {48}},\ \bibinfo {pages} {212} (\bibinfo {year}
  {2008})}\BibitemShut {NoStop}%
\bibitem [{\citenamefont {Scott}(2003)}]{scott2003computation}%
  \BibitemOpen
  \bibfield  {author} {\bibinfo {author} {\bibfnamefont {B.~D.}\ \bibnamefont
  {Scott}},\ }\href
  {http://iopscience.iop.org/article/10.1088/0741-3335/45/12A/025/meta}
  {\bibfield  {journal} {\bibinfo  {journal} {Plasma Phys. Controlled Fusion}\
  }\textbf {\bibinfo {volume} {45}},\ \bibinfo {pages} {A385} (\bibinfo {year}
  {2003})}\BibitemShut {NoStop}%
\bibitem [{\citenamefont {Scott}\ \emph {et~al.}(2010)\citenamefont {Scott},
  \citenamefont {Kendl},\ and\ \citenamefont {Ribeiro}}]{scott2010nonlinear}%
  \BibitemOpen
  \bibfield  {author} {\bibinfo {author} {\bibfnamefont {B.~D.}\ \bibnamefont
  {Scott}}, \bibinfo {author} {\bibfnamefont {A.}~\bibnamefont {Kendl}}, \ and\
  \bibinfo {author} {\bibfnamefont {T.}~\bibnamefont {Ribeiro}},\ }\href
  {https://onlinelibrary.wiley.com/doi/abs/10.1002/ctpp.201010039} {\bibfield
  {journal} {\bibinfo  {journal} {Contrib. Plasma Phys.}\ }\textbf {\bibinfo
  {volume} {50}},\ \bibinfo {pages} {228} (\bibinfo {year} {2010})}\BibitemShut
  {NoStop}%
\bibitem [{\citenamefont {Shi}\ \emph {et~al.}(2015)\citenamefont {Shi},
  \citenamefont {Hakim},\ and\ \citenamefont {Hammett}}]{shi2015gyrokinetic}%
  \BibitemOpen
  \bibfield  {author} {\bibinfo {author} {\bibfnamefont {E.}~\bibnamefont
  {Shi}}, \bibinfo {author} {\bibfnamefont {A.}~\bibnamefont {Hakim}}, \ and\
  \bibinfo {author} {\bibfnamefont {G.}~\bibnamefont {Hammett}},\ }\href
  {https://doi.org/10.1063/1.4907160} {\bibfield  {journal} {\bibinfo
  {journal} {Phys. Plasmas}\ }\textbf {\bibinfo {volume} {22}},\ \bibinfo
  {pages} {022504} (\bibinfo {year} {2015})}\BibitemShut {NoStop}%
\bibitem [{\citenamefont {Shi}\ \emph {et~al.}(2017)\citenamefont {Shi},
  \citenamefont {Hammett}, \citenamefont {Stoltzfus-Dueck},\ and\ \citenamefont
  {Hakim}}]{shi2017gyrokinetic}%
  \BibitemOpen
  \bibfield  {author} {\bibinfo {author} {\bibfnamefont {E.}~\bibnamefont
  {Shi}}, \bibinfo {author} {\bibfnamefont {G.}~\bibnamefont {Hammett}},
  \bibinfo {author} {\bibfnamefont {T.}~\bibnamefont {Stoltzfus-Dueck}}, \ and\
  \bibinfo {author} {\bibfnamefont {A.}~\bibnamefont {Hakim}},\ }\href
  {https://doi.org/10.1017/S002237781700037X} {\bibfield  {journal} {\bibinfo
  {journal} {J. Plasma Phys.}\ }\textbf {\bibinfo {volume} {83}},\ \bibinfo
  {pages} {905830304} (\bibinfo {year} {2017})}\BibitemShut {NoStop}%
\bibitem [{\citenamefont {Shi}(2017)}]{Shi2017thesis}%
  \BibitemOpen
  \bibfield  {author} {\bibinfo {author} {\bibfnamefont {E.}~\bibnamefont
  {Shi}},\ }\emph {\bibinfo {title} {Gyrokinetic Continuum Simulation of
  Turbulence in Open-Field-Line\ Plasmas}},\ \href
  {https://arxiv.org/abs/1708.07283} {Ph.D. thesis},\ \bibinfo  {school}
  {Princeton University} (\bibinfo {year} {2017})\BibitemShut {NoStop}%
\bibitem [{\citenamefont {Shi}\ \emph {et~al.}(2019)\citenamefont {Shi},
  \citenamefont {Hammett}, \citenamefont {Stoltzfus-Dueck},\ and\ \citenamefont
  {Hakim}}]{shi2019full}%
  \BibitemOpen
  \bibfield  {author} {\bibinfo {author} {\bibfnamefont {E.~L.}\ \bibnamefont
  {Shi}}, \bibinfo {author} {\bibfnamefont {G.~W.}\ \bibnamefont {Hammett}},
  \bibinfo {author} {\bibfnamefont {T.}~\bibnamefont {Stoltzfus-Dueck}}, \ and\
  \bibinfo {author} {\bibfnamefont {A.}~\bibnamefont {Hakim}},\ }\href
  {https://doi.org/10.1063/1.5074179} {\bibfield  {journal} {\bibinfo
  {journal} {Physics of Plasmas}\ }\textbf {\bibinfo {volume} {26}},\ \bibinfo
  {pages} {012307} (\bibinfo {year} {2019})}\BibitemShut {NoStop}%
\bibitem [{\citenamefont {Zeiler}\ \emph {et~al.}(1997)\citenamefont {Zeiler},
  \citenamefont {Drake},\ and\ \citenamefont {Rogers}}]{zeiler1997nonlinear}%
  \BibitemOpen
  \bibfield  {author} {\bibinfo {author} {\bibfnamefont {A.}~\bibnamefont
  {Zeiler}}, \bibinfo {author} {\bibfnamefont {J.}~\bibnamefont {Drake}}, \
  and\ \bibinfo {author} {\bibfnamefont {B.}~\bibnamefont {Rogers}},\ }\href
  {https://doi.org/10.1063/1.872368} {\bibfield  {journal} {\bibinfo  {journal}
  {Physics of Plasmas}\ }\textbf {\bibinfo {volume} {4}},\ \bibinfo {pages}
  {2134} (\bibinfo {year} {1997})}\BibitemShut {NoStop}%
\bibitem [{\citenamefont {Liu}\ and\ \citenamefont {Shu}(2000)}]{liu2000high}%
  \BibitemOpen
  \bibfield  {author} {\bibinfo {author} {\bibfnamefont {J.-G.}\ \bibnamefont
  {Liu}}\ and\ \bibinfo {author} {\bibfnamefont {C.-W.}\ \bibnamefont {Shu}},\
  }\href {https://www.sciencedirect.com/science/article/pii/S0021999100964751}
  {\bibfield  {journal} {\bibinfo  {journal} {J. Comput. Phys.}\ }\textbf
  {\bibinfo {volume} {160}},\ \bibinfo {pages} {577} (\bibinfo {year}
  {2000})}\BibitemShut {NoStop}%
\bibitem [{\citenamefont {Gottlieb}\ \emph {et~al.}(2001)\citenamefont
  {Gottlieb}, \citenamefont {Shu},\ and\ \citenamefont
  {Tadmor}}]{gottlieb2001strong}%
  \BibitemOpen
  \bibfield  {author} {\bibinfo {author} {\bibfnamefont {S.}~\bibnamefont
  {Gottlieb}}, \bibinfo {author} {\bibfnamefont {C.-W.}\ \bibnamefont {Shu}}, \
  and\ \bibinfo {author} {\bibfnamefont {E.}~\bibnamefont {Tadmor}},\ }\href
  {https://epubs.siam.org/doi/abs/10.1137/S003614450036757X} {\bibfield
  {journal} {\bibinfo  {journal} {SIAM Rev.}\ }\textbf {\bibinfo {volume}
  {43}},\ \bibinfo {pages} {89} (\bibinfo {year} {2001})}\BibitemShut {NoStop}%
\bibitem [{\citenamefont {Beer}\ \emph {et~al.}(1995)\citenamefont {Beer},
  \citenamefont {Cowley},\ and\ \citenamefont {Hammett}}]{beer1995field}%
  \BibitemOpen
  \bibfield  {author} {\bibinfo {author} {\bibfnamefont {M.~A.}\ \bibnamefont
  {Beer}}, \bibinfo {author} {\bibfnamefont {S.~C.}\ \bibnamefont {Cowley}}, \
  and\ \bibinfo {author} {\bibfnamefont {G.}~\bibnamefont {Hammett}},\ }\href
  {https://aip.scitation.org/doi/10.1063/1.871232} {\bibfield  {journal}
  {\bibinfo  {journal} {Phys. Plasmas}\ }\textbf {\bibinfo {volume} {2}},\
  \bibinfo {pages} {2687} (\bibinfo {year} {1995})}\BibitemShut {NoStop}%
\bibitem [{\citenamefont {Hammett}\ \emph {et~al.}(1993)\citenamefont
  {Hammett}, \citenamefont {Beer}, \citenamefont {Dorland}, \citenamefont
  {Cowley},\ and\ \citenamefont {Smith}}]{hammett1993developments}%
  \BibitemOpen
  \bibfield  {author} {\bibinfo {author} {\bibfnamefont {G.}~\bibnamefont
  {Hammett}}, \bibinfo {author} {\bibfnamefont {M.}~\bibnamefont {Beer}},
  \bibinfo {author} {\bibfnamefont {W.}~\bibnamefont {Dorland}}, \bibinfo
  {author} {\bibfnamefont {S.}~\bibnamefont {Cowley}}, \ and\ \bibinfo {author}
  {\bibfnamefont {S.}~\bibnamefont {Smith}},\ }\href
  {http://iopscience.iop.org/article/10.1088/0741-3335/35/8/006/meta}
  {\bibfield  {journal} {\bibinfo  {journal} {Plasma Phys. Controlled Fusion}\
  }\textbf {\bibinfo {volume} {35}},\ \bibinfo {pages} {973} (\bibinfo {year}
  {1993})}\BibitemShut {NoStop}%
\bibitem [{\citenamefont {Scott}(1998)}]{scott1998global}%
  \BibitemOpen
  \bibfield  {author} {\bibinfo {author} {\bibfnamefont {B.}~\bibnamefont
  {Scott}},\ }\href {https://aip.scitation.org/doi/10.1063/1.872907} {\bibfield
   {journal} {\bibinfo  {journal} {Phys. Plasmas}\ }\textbf {\bibinfo {volume}
  {5}},\ \bibinfo {pages} {2334} (\bibinfo {year} {1998})}\BibitemShut
  {NoStop}%
\bibitem [{\citenamefont {Mavrin}(2017)}]{mavrin2017radiative}%
  \BibitemOpen
  \bibfield  {author} {\bibinfo {author} {\bibfnamefont {A.}~\bibnamefont
  {Mavrin}},\ }\href
  {https://link.springer.com/article/10.1007/s10894-017-0136-z} {\bibfield
  {journal} {\bibinfo  {journal} {J. Fusion Energ.}\ }\textbf {\bibinfo
  {volume} {36}},\ \bibinfo {pages} {161} (\bibinfo {year} {2017})}\BibitemShut
  {NoStop}%
\bibitem [{\citenamefont {Li}\ \emph {et~al.}(2017)\citenamefont {Li},
  \citenamefont {Wang}, \citenamefont {Sun}, \citenamefont {Meng},
  \citenamefont {Zhou},\ and\ \citenamefont {Liu}}]{li2017edge}%
  \BibitemOpen
  \bibfield  {author} {\bibinfo {author} {\bibfnamefont {B.}~\bibnamefont
  {Li}}, \bibinfo {author} {\bibfnamefont {X.}~\bibnamefont {Wang}}, \bibinfo
  {author} {\bibfnamefont {C.}~\bibnamefont {Sun}}, \bibinfo {author}
  {\bibfnamefont {C.}~\bibnamefont {Meng}}, \bibinfo {author} {\bibfnamefont
  {A.}~\bibnamefont {Zhou}}, \ and\ \bibinfo {author} {\bibfnamefont
  {D.}~\bibnamefont {Liu}},\ }\href {https://doi.org/10.1063/1.4983624}
  {\bibfield  {journal} {\bibinfo  {journal} {Phys. Plasmas}\ }\textbf
  {\bibinfo {volume} {24}},\ \bibinfo {pages} {055905} (\bibinfo {year}
  {2017})}\BibitemShut {NoStop}%
\bibitem [{\citenamefont {Francisquez}\ \emph {et~al.}(2017)\citenamefont
  {Francisquez}, \citenamefont {Zhu},\ and\ \citenamefont
  {Rogers}}]{francisquez2017global}%
  \BibitemOpen
  \bibfield  {author} {\bibinfo {author} {\bibfnamefont {M.}~\bibnamefont
  {Francisquez}}, \bibinfo {author} {\bibfnamefont {B.}~\bibnamefont {Zhu}}, \
  and\ \bibinfo {author} {\bibfnamefont {B.}~\bibnamefont {Rogers}},\ }\href
  {https://doi.org/10.1088/1741-4326/aa7f23} {\bibfield  {journal} {\bibinfo
  {journal} {Nucl. Fusion}\ }\textbf {\bibinfo {volume} {57}},\ \bibinfo
  {pages} {116049} (\bibinfo {year} {2017})}\BibitemShut {NoStop}%
\bibitem [{\citenamefont {Dudson}\ \emph {et~al.}(2016)\citenamefont {Dudson},
  \citenamefont {Madsen}, \citenamefont {Omotani}, \citenamefont {Hill},
  \citenamefont {Easy},\ and\ \citenamefont
  {L{\o}iten}}]{dudson2016verification}%
  \BibitemOpen
  \bibfield  {author} {\bibinfo {author} {\bibfnamefont {B.~D.}\ \bibnamefont
  {Dudson}}, \bibinfo {author} {\bibfnamefont {J.}~\bibnamefont {Madsen}},
  \bibinfo {author} {\bibfnamefont {J.}~\bibnamefont {Omotani}}, \bibinfo
  {author} {\bibfnamefont {P.}~\bibnamefont {Hill}}, \bibinfo {author}
  {\bibfnamefont {L.}~\bibnamefont {Easy}}, \ and\ \bibinfo {author}
  {\bibfnamefont {M.}~\bibnamefont {L{\o}iten}},\ }\href
  {https://doi.org/10.1063/1.4953429} {\bibfield  {journal} {\bibinfo
  {journal} {Phys. Plasmas}\ }\textbf {\bibinfo {volume} {23}},\ \bibinfo
  {pages} {062303} (\bibinfo {year} {2016})}\BibitemShut {NoStop}%
\bibitem [{\citenamefont {Halpern}\ \emph {et~al.}(2016)\citenamefont
  {Halpern}, \citenamefont {Ricci}, \citenamefont {Jolliet}, \citenamefont
  {Loizu}, \citenamefont {Morales}, \citenamefont {Mosetto}, \citenamefont
  {Musil}, \citenamefont {Riva}, \citenamefont {Tran},\ and\ \citenamefont
  {Wersal}}]{halpern2016gbs}%
  \BibitemOpen
  \bibfield  {author} {\bibinfo {author} {\bibfnamefont {F.}~\bibnamefont
  {Halpern}}, \bibinfo {author} {\bibfnamefont {P.}~\bibnamefont {Ricci}},
  \bibinfo {author} {\bibfnamefont {S.}~\bibnamefont {Jolliet}}, \bibinfo
  {author} {\bibfnamefont {J.}~\bibnamefont {Loizu}}, \bibinfo {author}
  {\bibfnamefont {J.}~\bibnamefont {Morales}}, \bibinfo {author} {\bibfnamefont
  {A.}~\bibnamefont {Mosetto}}, \bibinfo {author} {\bibfnamefont
  {F.}~\bibnamefont {Musil}}, \bibinfo {author} {\bibfnamefont
  {F.}~\bibnamefont {Riva}}, \bibinfo {author} {\bibfnamefont {T.-M.}\
  \bibnamefont {Tran}}, \ and\ \bibinfo {author} {\bibfnamefont
  {C.}~\bibnamefont {Wersal}},\ }\href
  {https://doi.org/10.1016/j.jcp.2016.03.040} {\bibfield  {journal} {\bibinfo
  {journal} {J. Comput. Phys.}\ }\textbf {\bibinfo {volume} {315}},\ \bibinfo
  {pages} {388} (\bibinfo {year} {2016})}\BibitemShut {NoStop}%
\bibitem [{\citenamefont {Terry}(2000)}]{terry2000suppression}%
  \BibitemOpen
  \bibfield  {author} {\bibinfo {author} {\bibfnamefont {P.}~\bibnamefont
  {Terry}},\ }\href {https://doi.org/10.1103/RevModPhys.72.10} {\bibfield
  {journal} {\bibinfo  {journal} {Rev. Mod. Phys.}\ }\textbf {\bibinfo {volume}
  {72}},\ \bibinfo {pages} {109} (\bibinfo {year} {2000})}\BibitemShut
  {NoStop}%
\bibitem [{\citenamefont {D'Ippolito}\ \emph {et~al.}(2011)\citenamefont
  {D'Ippolito}, \citenamefont {Myra},\ and\ \citenamefont
  {Zweben}}]{d2011convective}%
  \BibitemOpen
  \bibfield  {author} {\bibinfo {author} {\bibfnamefont {D.}~\bibnamefont
  {D'Ippolito}}, \bibinfo {author} {\bibfnamefont {J.}~\bibnamefont {Myra}}, \
  and\ \bibinfo {author} {\bibfnamefont {S.}~\bibnamefont {Zweben}},\ }\href
  {https://doi.org/10.1063/1.3594609} {\bibfield  {journal} {\bibinfo
  {journal} {Phys. Plasmas}\ }\textbf {\bibinfo {volume} {18}},\ \bibinfo
  {pages} {060501} (\bibinfo {year} {2011})}\BibitemShut {NoStop}%
\bibitem [{\citenamefont {Pitts}\ \emph {et~al.}(2007)\citenamefont {Pitts},
  \citenamefont {Andrew}, \citenamefont {Arnoux}, \citenamefont {Eich},
  \citenamefont {Fundamenski}, \citenamefont {Huber}, \citenamefont {Silva},\
  and\ \citenamefont {Tskhakaya}}]{Pitts_2007}%
  \BibitemOpen
  \bibfield  {author} {\bibinfo {author} {\bibfnamefont {R.}~\bibnamefont
  {Pitts}}, \bibinfo {author} {\bibfnamefont {P.}~\bibnamefont {Andrew}},
  \bibinfo {author} {\bibfnamefont {G.}~\bibnamefont {Arnoux}}, \bibinfo
  {author} {\bibfnamefont {T.}~\bibnamefont {Eich}}, \bibinfo {author}
  {\bibfnamefont {W.}~\bibnamefont {Fundamenski}}, \bibinfo {author}
  {\bibfnamefont {A.}~\bibnamefont {Huber}}, \bibinfo {author} {\bibfnamefont
  {C.}~\bibnamefont {Silva}}, \ and\ \bibinfo {author} {\bibfnamefont
  {D.}~\bibnamefont {Tskhakaya}},\ }\href {\doibase
  10.1088/0029-5515/47/11/005} {\bibfield  {journal} {\bibinfo  {journal}
  {Nuclear Fusion}\ }\textbf {\bibinfo {volume} {47}},\ \bibinfo {pages} {1437}
  (\bibinfo {year} {2007})}\BibitemShut {NoStop}%
\end{thebibliography}%

\end{document}